\documentclass[12pt,preprint]{aastex}

\usepackage{epstopdf}

\slugcomment{Accepted by ApJ: July 26, 2011}

\shorttitle{Laboratory spectra of astrophysically-relevant glasses}
\shortauthors{Speck et al.}

\begin{document}
\title{Disordered Silicates in Space: 
a Study of Laboratory Spectra of ``Amorphous'' Silicates}

\author{Angela K.\ Speck\altaffilmark{1,2},
Alan G. Whittington\altaffilmark{2}
Anne. M. Hofmeister\altaffilmark{3}}

\altaffiltext{1}{Department of Physics and Astronomy, University of Missouri, 
Columbia, MO 65211}
\altaffiltext{2}{Department of Geological Sciences, University of Missouri, 
Columbia, MO 65211}
\altaffiltext{3}{Department of Earth and Planetary Sciences, Washington 
University, St. Louis, MO 63130, USA}

\begin{abstract}

We present a laboratory study of silicate glasses of 
astrophysically relevant compositions 
including olivines, pyroxenes and melilites.
With emphasis on the classic Si--O stretching feature near 10$\mu$m, 
we compare infrared spectra of our new samples with laboratory spectra
on ostensibly similar compositions, and also with synthetic silicate 
spectral data commonly used in dust modeling.
Several different factors affect spectral features including sample 
chemistry 
(e.g., polymerization, Mg/Fe ratio, oxidation state and Al-content) 
and different sample preparation techniques lead to variations 
in porosity, density and water content. The convolution of 
chemical and physical effects makes it difficult to attribute changes in 
spectral parameters to any given variable.
It is important that detailed chemical and structural characterization be 
provided along with laboratory spectra.
In addition to composition and density, we measured the glass transition temperatures for the samples 
which place upper limits on the formation/processing temperatures of these 
solids in space.

Popular synthetically-generated optical functions 
do not have spectral features that match any of our glass samples.
However, the $\sim$10$\mu$m feature generated by the 
synthetic data rarely exactly matches the shape and peak position of 
astronomically observed silicate features. 
Our comparison with the synthetic spectra allows astronomers to determine 
likely candidates amongst our glass samples for matching astronomical 
observations.

\end{abstract}


\keywords{stars: AGB and post-AGB
--- (stars:) circumstellar matter 
--- (ISM:) dust, extinction
--- infrared: stars  }




\section{Introduction}
\label{intro}

Understanding the nature and formation of cosmic dust is crucial to our 
understanding of the cosmos. Over its 50-year history, infrared (IR) 
astronomy has shown 
that dust contributes to the physical processes inherent in star 
formation and mass-loss from evolved stars, as well as to several interstellar 
processes such as gas heating and the formation of molecules
\citep[e.g.][]{vk02,draine03,krishna05,krugel08}.
In particular, silicate grains dominate dust emission in many astrophysical 
environments. The ``amorphous'' $\sim10\mu$m and $\sim18\mu$m silicate 
spectral features have been observed in almost every direction and to almost 
any distance, but the precise nature of this silicate dust remains a mystery.
%
%
Here we present a laboratory investigation of  amorphous silicates, the 
type of dust grains most frequently inferred to exist from observational data.

\subsection{A brief history of the ``amorphous'' silicate spectral features}
\label{history}

The classic ``10\,$\mu$m'' silicate feature was first observed in the late 
sixties in the IR spectra of several M-type giants and 
red supergiants \citep[RSGs;][]{gillett68}.
Shortly thereafter a 10\,$\mu$m absorption feature was discovered in the 
interstellar medium \citep[ISM;][]{knacke69,hackwell70}.
Since then, it has been found to be almost ubiquitous, occurring in many 
astrophysical environments including the solar system and extrasolar planetary 
systems \citep[e.g.,][and references therein]{mann06}, the circumstellar 
regions of both young stellar objects and evolved intermediate mass stars 
\citep[asymptotic giant branch; AGB stars, and planetary nebulae; e.g.,][]{speck00,casassus01}; many lines of sight through the interstellar 
medium in our own galaxy \citep[e.g.,][]{chiar07}; and
in nearby and distant galaxies \citep[e.g.,][]{hao05}. 
Initially this feature was attributed to silicate minerals \citep{woolf69}, 
based on mixtures of spectra of crystalline silicate species predicted to 
form by theoretical models \citep{gaustad63,gilman69}. However, laboratory 
spectra of crystalline silicate minerals showed more structure within the 
feature than observed in the astronomical spectra 
\citep[see e.g.,][]{woolf73,huffman73}. 
Subsequent comparison with natural glasses \citep[obsidian and basaltic glass; 
from e.g.,][]{pollack73} and with artificially disordered silicates 
\citep{day79,kh79} 
showed that ``amorphous'' silicate was a better candidate 
for the 10\,$\mu$m feature than any individual crystalline silicate mineral. 
Since then, it has been commonly assumed that description of silicate as 
disordered or ``amorphous'' is synonymous with glassy silicate. 
%
However, this is an 
oversimplification. The term ``glass'' has a specific definition, i.e. the 
solid has no long-range order beyond nearest-neighbor atoms. 
``Crystalline'' is often taken to mean single crystals, but it is possible to 
form poly-nanocrystalline agglomerates, with a continuum which essentially 
extends from a true glass to a single crystal grain. Furthermore, natural and 
synthetic ``glasses'' often contain microlites\footnote{%
micro- or nano-crystalline inclusions within a glassy matrix
\citep[see e.g.,][]{pollack73,jager94}. 
Prior to \citet{zachariasen} the difference between nanocrystalline 
(i.e. ceramic) and glassy solids was not understood.}%
In addition, one might expect agglomerated particles to be polymineralic, and 
possibly to contain both crystalline and glassy constituents. 
\citet{NuthHecht} introduced the idea of 
``chaotic silicate'' in which the level of disorder is even greater than for 
glass. A chaotic silicate does not have to be stoichiometric, 
can contain different compositional zones within a 
single grain, 
and may be porous, and therefore much lower density than a glass. 
This range of 
possible grain types is demonstrated schematically in Fig.~\ref{fig:XtalC}.

In astrophysical environments, whether a solid is glassy, crystalline or some 
combination of the two has implications for its 
formation, and subsequent processing, evolution and destruction, and thus it 
is important to have tools to distinguish between such grain types. 
For example, true glasses with no inclusions will not transform into crystals 
below their 
glass transition temperature ($T_g$; see \S~\ref{silglass}), 
whereas a glass already containing microlites can continue to crystallize at 
slightly lower 
temperatures. 
In the case of terrestrial obsidians, with 
900\,K $ < T_g < $ 1000\,K, elemental diffusion profiles suggest that 
crystal growth can continue down to $\sim$700\,K  \citep[e.g.,][]{watkins09}. 
If poly-nano-crystalline grains can be distinguished spectroscopically in the 
lab from truly glassy grains, we can test for their presence in astrophysical 
environments. For instance \citet{SH04} showed that there is 
a difference between single crystal and polynanocrystalline silicon carbide 
(SiC), while 
\citet{STH05} showed that glassy SiC looks different than various  crystalline 
samples. These laboratory data were invoked to explain changes in SiC grains 
formed as carbon stars evolve.
For silicates, laboratory data on poly-nano-crystalline samples are lacking.
Whether grains are glassy or poly-nano-crystalline is an indicator 
of whether
grains form or are processed above 
$T_g$. If the two forms cannot be distinguished in laboratory spectra, then 
the ``amorphous'' nature of silicates in space would no longer necessarily 
imply a truly glassy structure, allowing the possibility of higher dust 
formation temperatures.

Many observations have shown that the 10\,$\mu$m silicate feature varies from 
object to object and even within a single object both temporally 
\citep[e.g.,][]{monnier98} and spatially (e.g. in $\eta$ Car, N. Smith, 
Pers. Comm). Within a single type of astrophysical object, the feature shows 
huge variations in terms of its peak position, width and its ratio to the 
$\sim 18\mu$m feature \citep*[e.g.][]{speck98,ohm92}. 
Variations in feature shape from star to star cannot be explained in terms 
of optical depth or grain temperature effects. 
Several interpretations of these observations have been suggested including:
grain size effect  \citep[e.g.,][]{papoular83};
Mg/Fe ratio and (Mg+Fe)/Si ratio \citep[e.g.][]{dorschner95}; 
inclusion of oxide grains \citep[e.g.,][]{speck00}; 
increasing crystallinity \citep[e.g.,][]{sylvester98,bouwman01}; 
grain shape \citep[e.g.,][]{min07}; and 
grain porosity \citep[e.g.,][]{vh08,henning93}.
However, all models of these effects utilize laboratory spectra, 
and are only as reliable as the data that goes into them.

The near ubiquity of  ``amorphous'' silicate features and their variations 
in strength, width, peak position and the ratio of the 10\,$\mu$m/18\,$\mu$m 
features potentially provide the diagnostic tools to understand the detailed 
mechanisms by which dust is formed, processed and destroyed. However, 
existing laboratory and synthetic spectra are not sufficiently well 
understood to achieve this goal.

\subsection{A brief synopsis of existing laboratory and synthetic spectra for disordered silicates}
\label{prevlab}

Since the discovery of the 10\,$\mu$m feature, there have been many laboratory 
studies producing IR spectra and optical functions\footnote{Usually these are called optical ``constants'', but they are wavelength-dependent quantities, so we prefer ``functions''} of various samples for 
comparison with and modeling of observational data. In addition, synthetic 
optical functions
 have been derived from observational spectra, often combined with laboratory mineral data, 
in order to match the observed features 
\citep*[e.g.,][]{dl84,vk88,ohm92}.

The first laboratory spectra used in astronomical silicate studies were of 
crystalline silicates and natural glasses \citep[obsidian and basaltic glass, 
which contain some microlites; e.g.,][]{pollack73}. 
Various studies produced ``amorphous'' samples through 
chemical vapor deposition \citep[e.g.,][]{day79}, smokes 
\citep[e.g.,][]{nd82}, ion-irradiation of crystalline samples 
\citep[e.g.,][]{kh79},  laser ablation of crystalline samples 
\citep{sd96}, and quenching melts to glass 
\citep[e.g.,][]{dorschner95,jager03}. 
However, these techniques  yield different results.
For instance the peak positions, full width at half maxima (FHWMa) and ratios of 
the strengths of the 10 and 18\,$\mu$m features vary between datasets 
even though the materials investigated are ostenibly the same composition.
Spectra from samples with the same reported composition vary, 
which should not occur if the samples had the same structures, and if all spectra were obtained under optically thin conditions.
%
%
%
A detailed comparison between existing and new laboratory data are given in 
\S~\ref{comparison}.
The sample preparation techniques vary widely and lead to a range of 
disordered structures. Unfortunately we do not have 
sufficient information on the physical structure or chemical characteristics 
of previously studied samples to 
determine the effects quantitively, but a qualitative analysis presented here 
highlights the need to make such sample information available.

The data used most commonly for modeling astronomical environments are  
synthetic optical functions such as \citet*[hereafter DL]{dl84} and 
\citet*[hereafter OHM]{ohm92}. 
These data are favored over laboratory spectra
because they  have broad wavelength coverage, which is not true 
of individual laboratory datasets. However, these synthetic functions were produced using 
compilations of laboratory spectral data and astronomical observed dust opacities from which new 
optical functions were calculated. The derived optical functions were then modified specifically to match 
the astronomical 
observations. 
For instance, in the \citeauthor{dl84} data, the 9.7\,$\mu$m feature
is entirely derived from observational opacities, 
while the NIR-NUV section of their optical function uses laboratory data from 
crystalline olivine studies, and the FUV/X-ray region uses laboratory data 
for crystalline alumina (Al$_2$O$_3$). Both \citeauthor{dl84} and \citeauthor{ohm92} blend
laboratory and astronomical data and their optical functions will match some spectral features, and can be used for 
comparison of optical depths between different dusty environments.
However, they do not 
represent real solids and thus cannot be used to determine the true nature of 
dust in space, how it varies spatially or temporally, or why.

\subsection{A brief guide to the structure of silicate minerals, glasses and the glass transition}
\label{silglass}

The basic building block of silicates is the SiO$_{4}^{4-}$ tetrahedron. 
These can be linked in a framework, with each oxygen shared between two 
tetrahedra (e.g. SiO$_2$ minerals and feldspars), or they can be linked in 
chains (e.g. pyroxenes such as 
diopside [Di; CaMgSi$_2$O$_6$], 
enstatite [En;MgSiO$_3$]) or they can be isolated tetrahedra 
(e.g. the olivines series: forsterite [fo; Mg$_2$SiO$_4$] to 
fayalite [fa; Fe$_2$SiO$_4$]). 
In all cases, the non-shared oxygens (known as non-bridging oxygens or NBOs) 
are charge-balanced by other cations 
(e.g. Mg$^{2+}$, Fe$^{2+}$, Ca$^{2+}$, Na$^+$, etc). 
The Ca-Mg-Al silicates that are expected to 
form in circumstellar environments are dominantly of pyroxene and olivine 
composition.

Within each mineral group, solid solution allows compositions to vary
 between end-members. 
In the case of similarly-sized cations the solution may be continuous
e.g., enstatite to ferrosilite (fs; FeSiO$_3$) and the olivine series.
Given the availability of other cations,  
end-member pyroxenes rarely occur in terrestrial or meteoritic samples. 
Where the substituting cation is a different size, solid solution is more limited,
for example between enstatite 
and diopside. In such cases, solid solution becomes less extensive at lower temperatures, 
so that cooling 
may lead to ``exsolution'' of two different pyroxene compositions from an 
initially homogeneous high-temperature solid. 

Minerals occur as crystals 
that possess long-range order, with very narrow distributions of bond angles 
and lengths. This leads both to anisotropy (properties varying with crystal 
orientation) and narrow spectral features. 
Silicate glasses are the ``frozen'' structural equivalents of liquids, 
possessing short-range order (so that local charge-balance is conserved for 
example) but lacking the long-range order that gives rise to symmetry and 
anisotropy in crystals. 
Glasses are therefore both isotropic and have broad spectral features.

\begin{figure}[t]
\includegraphics[angle=0,scale=0.8]{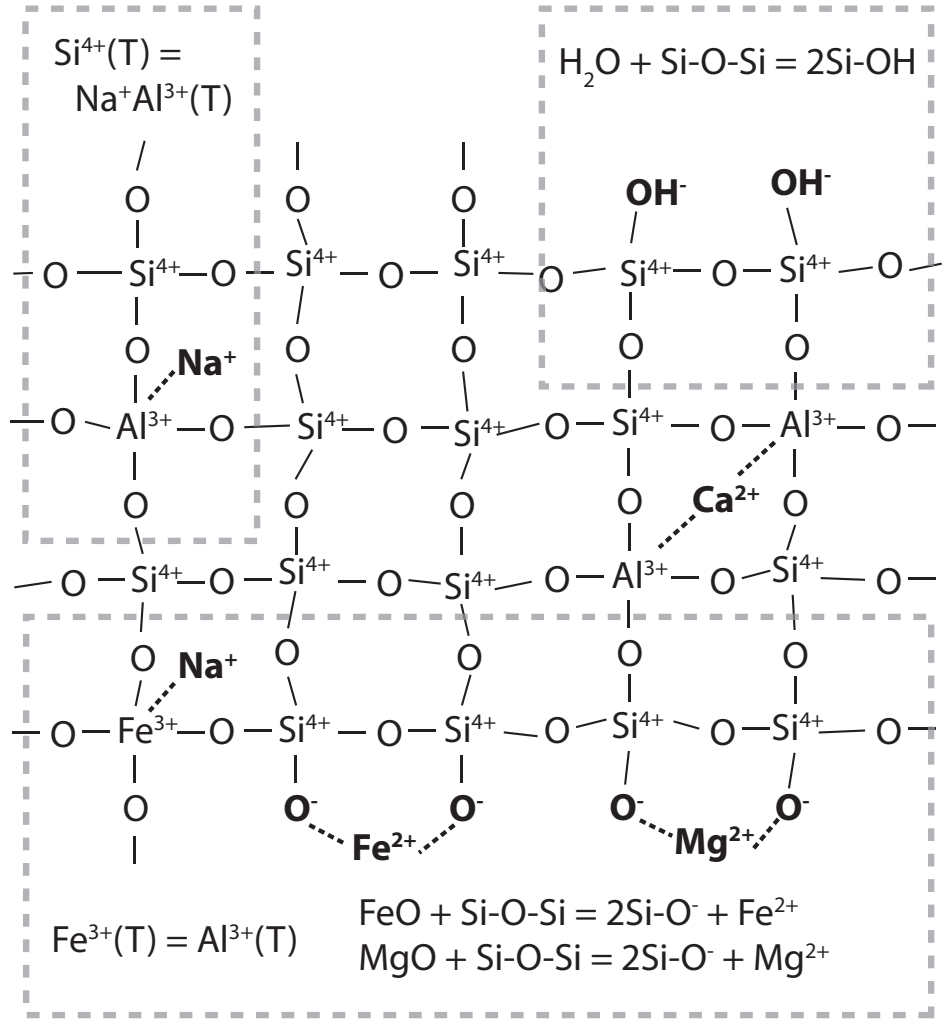}
\caption{\label{cartoon} 
Schematic structure of silicate glasses showing bridging oxygens joining 
tetrahedral units, and non-bridging oxygens coordinated by network-modifying 
cations. Trivalent tetrahedral cations (Al$^{3+}$, Fe$^{3+}$) 
must be charge-balanced by metal cations occupying interstices in the 
structure (e.g. Na$^{+}$, Ca$^{2+}$). The 
real three-dimensional structure of glass is a modified random network 
\citep[e.g.][]{henderson}.}
\end{figure}

The basic structural unit of silicate glasses and melts is the SiO$_{4}^{4-}$ 
tetrahedron, as is the case in crystalline silicate minerals. Oxygens linking 
tetrahedra are known as bridging oxygens (BO), while non-bridging oxygens 
(NBO) are coordinated by metal cations, which are termed network-modifiers in 
this role. Tetrahedral cations (T) include not only Si$^{4+}$, but also trivalent 
cations such as Al$^{3+}$ and Fe$^{3+}$; these must be charge-balanced by 
other cations (usually alkalis or alkaline earths; Fig.~\ref{cartoon}). 
The degree of polymerization of a melt or glass can be summarized by the ratio 
of NBO/T, which can range from 0 (fully polymerized, e.g. SiO$_2$) to 4 
(fully depolymerized, e.g. Mg$_2$SiO$_4$). In general, more polymerized melts 
are more viscous and have higher glass transition temperatures. 
On quenching a melt, its structure is ``frozen in'' at the glass transition if 
cooling is rapid enough to prevent crystallization. 
The glass transition is actually an interval, often approximated by the glass 
transition temperature ($T_g$) which is usually taken to be the temperature at 
which the viscosity is 10$^{12}$ Pa.s ($T_{12}$). 
Rapid cooling from a given temperature preserves the network present in the
liquid at that particular temperature. Because of this behavior, glasses of any
given composition can have subtle differences in structure that depend on
cooling rates.
The
temperature at which the glass has the same structure as the melt is called 
the fictive temperature ($T_f$).
%
See \citet{mysen05} for a comprehensive review of melt 
structure and properties.

Two issues whose importance will be discussed in the current work
are the role of water and the 
oxidation state of iron. At low water contents (less than about 1 wt.\% total 
H$_2$O), water dissolves in silicate glasses almost exclusively as hydroxyl (OH$^-$) 
ions \citep{stolper82}, and acts as a network modifier (Fig.~\ref{cartoon}). 
Compared to other modifier oxides such as Na$_2$O and MgO, water has a more 
dramatic effect in reducing melt viscosity and glass transition temperatures 
\citep{dingwell96}. Iron can play the role of network modifier 
(octahedral Fe$^{2+}$ or Fe$^{3+}$) or network-forming cation 
(tetrahedral Fe$^{3+}$). 
Consequently, the oxidation state of an iron-bearing glass or melt has a 
significant effect on its structure and properties.
%


%
From the perspective of cosmic dust formation, the glass transition 
temperature is essentially the temperature above which a given composition 
should form as or convert to crystalline solids, whereas solids formed below 
this 
temperature will be glassy if they cool sufficiently rapidly 
\citep{richet1993,swt08}.
More depolymerized melts are more difficult to quench, and melts less 
polymerized than pyroxenes (NBO/T $>$ 2) typically require extreme quench rates 
(100s K\,s$^{-1}$) 
using methods such as containerless laser processing to achieve truly glassy 
samples \citep{tangeman01a}.
For a given composition, faster cooling rates result in a higher T$_g$. 
This dependence can be determined by differential scanning calorimetry using 
different heating and cooling rates, and is used to determine the cooling rate 
of natural lava samples \citep{wilding95}. 

For depolymerized silicates (e.g. olivines and pyroxenes), if glassy grains 
form they must do so below their $T_g$ because the cooling rate required for 
quenching to a glass is extremely rapid. Highly polymerized silicates (e.g. 
silica, obsidian) can be cooled more slowly, over hours or days, and not 
crystallize \citep{bowen}.
However, the cooling timescales (months) determined by \citet{swt08} for 
AGB star 
circumstellar shells  are sufficiently long as to preclude the preservation of 
glassy/chaotic solids that form above $T_g$, because annealing timescales
are shorter than those for cooling for all but the most polymerized 
silicates. 

\subsection{The need for new data}

Modeling of silicate dust in space has been limited by the available laboratory data
The influence of 
various model parameters was investigated by \citet{jm76}, who found that 
using so-called ``clean'' (i.e. pure magnesium) silicate grains to model the 
observed 9.7\,$\mu$m features did not yield a good fit due to the lack of 
absorption by these grains in the visible and near-IR. They also found that 
just mixing in more absorbing grains did not solve the problem.
This led to the suggestion that the grains responsible for the 9.7\,$\mu$m 
feature are ``dirty'' silicates, i.e. Mg-silicates with impurities introduced 
into the matrix giving more opacity in the optical and near-IR. 

It is  known that NBO/T (polymerization) affects the spectra of amorphous 
silicates such that the peak position of the 10\,$\mu$m feature shifts 
redwards as NBO/T increases (e.g. \citeauthor{ohm92}).
%
%
%
Aluminium (Al) is a network former and  consequently Al content 
strongly affects NBO/T. 
\citet{mutschke98} suggested that Al may be an important component of
silicates in space that could explain why previous laboratory spectroscopic 
studies failed to match observational data.
Other cations may be equally important. Ca$^{2+}$ and Fe$^{2+}$ both substitute for Mg, while
 Fe$^{3+}$ will substitute for tetrahedral site (e.g. Si$^4+$ or Al$^{3+}$). 
Therefore, the oxidation state is another important variable
in addition to elemental substitutions.

Existing laboratory spectral data for ``amorphous'' silicates 
were produced using samples that are not 
sufficiently well-characterized  
to allow astronomers to interpret their 
observations without ambiguity. Here we present new laboratory spectra of 
several silicate glasses of astronomical relevance, and
discuss compositional factors that influence their spectral features.
We compare these new data 
with those previously available for ``amorphous'' silicates
and discuss how these samples compare to successfully-applied synthetic optical functions.
We find that the synthetic spectra cannot be well matched by
the conventionally assumed glassy silicate composition and discuss
whether astrophysical silicates need to be truly glassy.


\section{Experimental Methods}
\label{exp}

\subsection{Sample selection}

\begin{figure}[t]
\includegraphics[angle=0,scale=0.7]{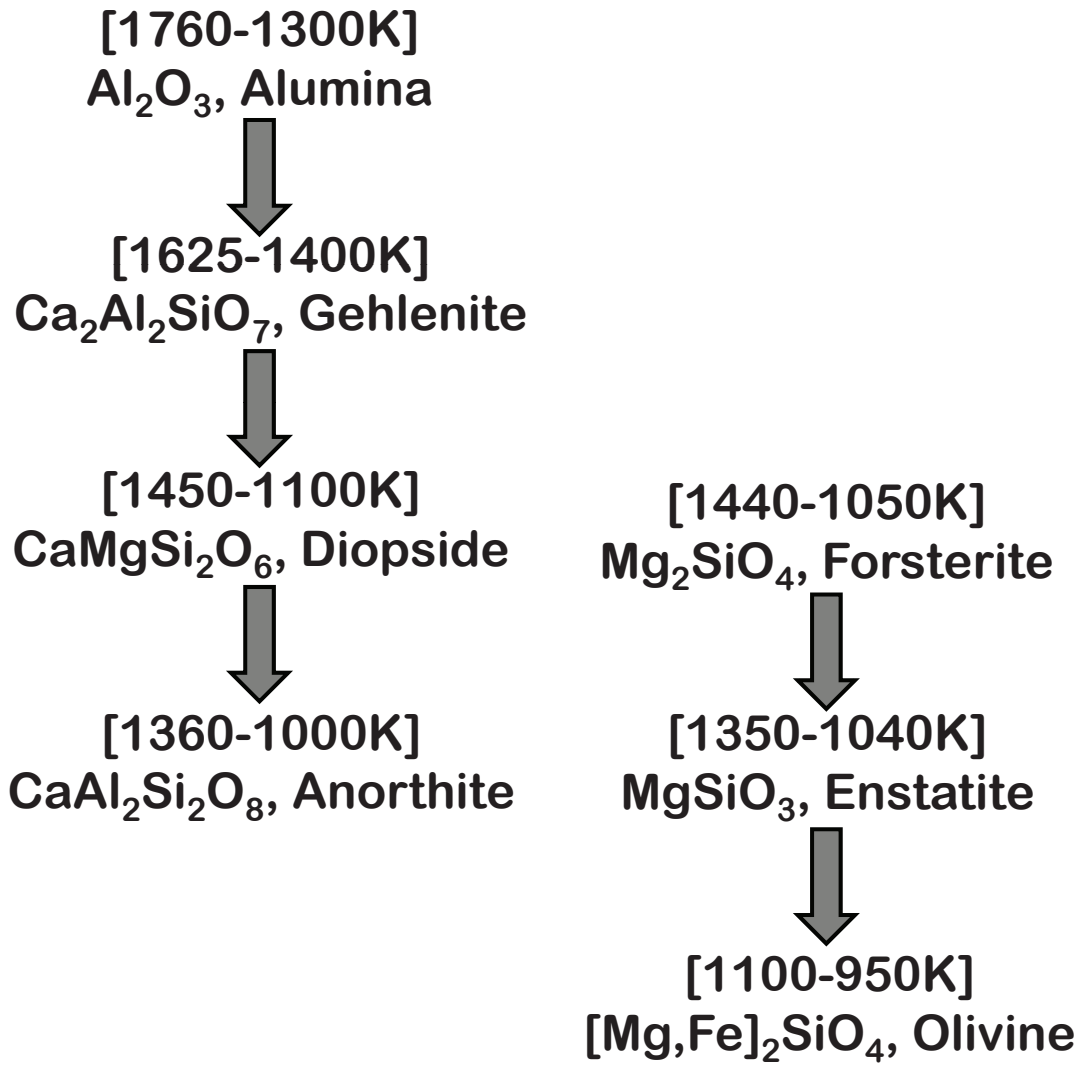}
\caption{\label{condseqfig} 
Predicted condensation sequence for O-rich environments \citep{grossman72}. }
\end{figure}

The bulk composition of silicate stardust lies somewhere between pyroxene 
(M$_2$Si$_2$O$_6$) and olivine (M$_2$SiO$_4$), where M indicates metal 
cations, with Mg and Fe being the most abundant. The cosmic Mg/Si ratio is 
predicted to be $\sim$1.02, while Fe/Si $\sim$0.84 \citep[e.g.,][]{lf99}.
However, most Fe is expected to combine with S, Ni, Cr, Mn and other 
siderophile elements into metallic grains \citep{gs99,lf99}.
This partitioning is reflected in the Earth, where most iron resides in the 
metallic core while the mantle is dominated by magnesium-rich silicate with 
Mg/Fe $\sim$9. The predicted bulk cosmic silicate would then be close to 
MgSiO$_3$, with minor amounts of iron leading to an atomic (Mg+Fe)/Si ratio 
slightly greater than 1. Determining the spectra of various olivine and 
pyroxene glasses is therefore of critical importance for
identifying the silicate mineralogy. 
The focus of this study is glass compositions 
for which data already exist in the 
astronomy literature i.e. predominantly olivines and pyroxenes 
(See Table~\ref{litdata} and references therein). 
Mg-rich endmembers are forsterite and enstatite, respectively.
Another mineral that is commonly discussed in astromineralogy is
diopside, which is also a pyroxene.
Diopside has been invoked to explain observed crystalline silicate features  
\citep[e.g.][]{Demyk2000,Kemper2002,Hony09,OO03} and appears in the 
classic condensation sequence for dust formation
\citep[e.g.][see Fig.~\ref{condseqfig}]{tielens90}. 
Furthermore, aluminous diopside formed in the experimental 
condensation study by \citet{Toppani2006}, and  \citet{Demyk2004} 
showed that crystalline diopside grains can easily be amorphized by 
heavy ion irradiation.


To complement the olivines and pyroxenes (Table~\ref{litdata})  we include four samples from the 
melilite series, whose endmembers are gehlenite (Ge; Ca$_2$Al$_2$SiO$_7$) and 
\r{a}kermanite (\r{A}k; Ca$_2$MgSi$_2$O$_7$).
gehlenite is predicted to be among the first-formed silicate grains 
\citep[e.g.,][see Figure~\ref{condseqfig}]{tielens90,lf99},
and the major repository for calcium and aluminum in dust, whereas pyroxenes 
are predicted to be among the most abundant grains, and the major repository 
for Mg and Si \citep[e.g.,][]{tielens90,lf99,gs99}. 
Furthermore aluminium and calcium are highly depleted from the gas phase and 
are assumed to be included in dust \citep{Whittet1992}.
Aluminum-rich silicates like gehlenite and other melilites are major constituents of 
Calcium-Aluminum-rich Inclusions (CAIs) and
gehlenite has been identified in red supergiants \citep{speck00} and 
Active Galactic Nuclei (AGN) environments 
\citep{jaffe04} and in meteorites \citep{stroud08,vollmer07}.
Furthermore, a recent Type Ib supernova (SN2005E) has been shown to contain more calcium 
than expected \citep{perets}. Indeed almost 50\% of its ejecta mass 
(or $>$0.1M$_\odot$) is attributed to calcium.
\citet{chihara07} reported laboratory spectra of crystalline melilites every 
10\% along the solid-solution join between \r{a}kermanite and gehlenite. 
\citet{mutschke98} reported laboratory spectra of two glasses: end-member 
gehlenite, and \r{a}k50-ge50. 
Given that the cosmic Mg/Ca ratio is $\sim$16, it seems likely that melilites 
could have substantial \r{a}kermanite contents. The difference in the 
structure of the glasses is profound: NBO/T is 0.67 for gehlenite and 3.0 for 
\r{a}kermanite, while Al/Si is 2 for gehlenite and 0 for \r{a}kermanite.

Both theoretical models \citep{lili} and 
observations \citep{jm76} suggest 
that some iron is incorporated into silicates.
\citet{jager94} presented spectra for two silicate glasses containing 
the most abundant dust forming elements (i.e. Mg, Si, Fe, Ca, Al, Na).
However, iron in their sample was partially oxidized (FeO/Fe$_2$O$_3 \sim$1) which
leads to problems in interpreting the spectrum (see \S~\ref{silglass}).
 With this in mind we present a sample we call ``Basalt'' 
($\rm Na_{0.09}Mg_{0.62}Ca_{0.69}Fe_{0.39}Ti_{0.10}Al_{0.06}Si_{2.16}O_6$),
which contains ferrous iron (Fe$^{2+}$).

While iron has been invoked to explain opacity problems, 
most iron is expected to combine with 
other siderophile elements to form metal or metal sulfide grains rather than 
silicate \citep{gs99,lf99}.
Consequently, we have synthesized an iron-free silicate glass using cosmic 
abundance ratios for 
Mg, Si, Al, Ca, and Na, yielding 
$\rm (Na_{0.10}Ca_{0.12}Mg_{1.86})(Al_{0.18}Si_{1.84})O_6 $. 
This sample does not include volatile elements  or iron and is quite close to 
enstatite in composition. 
This ``cosmic silicate''  is designed to test the \citet{stencel90} hypothesis 
that dust forms as chaotic solids 
with the elemental abundances of the gas.

\citet{pollack73} provided spectra of obsidian, a naturally occurring glassy 
silicate. Consequently we include obsidian glass in our sample, in part 
because this is the origin of the attribution of the $\sim$10\,$\mu$m feature 
to amorphous silicates. It is also useful for studying the effect of silicate 
structure and composition on spectral parameters because it is different from 
the more commonly assumed olivines and pyroxenes.

Finally, we include silica  (SiO$_2$) glass in our sample. Like obsidian,  the 
structure of SiO$_2$ is significantly different from olivines and pyroxenes 
and thus provides potential insight into the effects of structure on silicate 
spectral features. 
Furthermore, silica dust grains have been invoked to explain observed 
astronomical spectral features in both evolved stars \citep{speck00} and in 
young stellar objects \citep[e.g.][]{sargent06}.

The samples investigated here are listed in Table~\ref{tab:samples}.

\subsection{Sample synthesis and preparation}
\label{synth}

Samples designated with ``synthetic'' in their name are 
generated from mixtures of reagent-grade oxides and carbonates, 
providing glasses with compositions 
like those of end-member minerals. The ``cosmic'' silicate was also 
synthesized in this way.
In contrast, samples designated as ``remelt'' were generated by melting 
natural mineral samples. Consequently the ``remelt''-samples have compositions 
whose additional components reflect the impurities\footnote{
Unlike minerals, glasses do not have to have well-defined formulae. 
Some studies describe such non-stoichiometric glasses 
as having large quantities of impurities, but these 
non-mineral-end-member compositions are simply what the glass is made of.}
found in natural crystalline samples of the 
relevant minerals. Using both synthetic and remelted samples helps to 
demonstrate the effect and importance of small compositional variations 
in silicates.

Synthesis of silicate glasses from oxide and carbonate starting materials, 
was undertaken at the MU experimental petrology facility and the procedures are
described in detail by \citet{getson07}.
Melilite glasses were prepared by fusion in Pt crucibles and 
quenched by pouring into graphite molds (slow cooling), or on to a copper 
plate (faster cooling for less polymerized compositions). 
Glasses in the pyroxene series (including ``cosmic'' compositions) vary in 
their quenchability; diopside (CaMgSi$_2$O$_6$) is an excellent glass-former, 
while enstatite (MgSiO$_3$) crystallizes extremely rapidly. 

Glass of forsterite (Fo, Mg$_2$SiO$_4$) or olivine 
(Mg$_{2x}$Fe$_{2-2x}$SiO$_4$) can only be formed in 
the laboratory under special conditions.  Specifically, 
Fo glass in the form of 
50 to 200\,$\mu$m diameter beads was first produced by \citet{tangeman01a} at
Containerless Research by suspending small particles in argon gas and 
melting/quenching by pulsing with a laser.  
In addition to providing rapid cooling of $\sim$700\,K/sec, 
the lack of container promotes crystal-free glass formation.
%
Commercially prepared samples of Fo were purchased from 
Containerless Research, 
Inc.  
Iron-bearing olivine samples were not available.  
We were able to prepare ``basalt 
glass'' at Washington U. by flash melting rock on a glassy carbon substrate in 
a vacuum chamber with a CO$_2$ laser followed by rapid cooling. 
When applied to iron-bearing olivine or fayalite (Fe$_2$SiO$_4$) crystals
this approach failed to produce glass. 

\subsection{Chemical Analyses for Sample Composition Determination}

Samples were characterized by wavelength dispersive analysis (WDS) using 
standard procedures on the JEOL-733 and JXA-8200 electron microprobes at 
Washington University using ``Probe for Windows'' for data reduction 
(see http://www.probesoftware.com/). 
The measured data were corrected using CITZAF after \citet{Armstrong1995}. 
Oxide and silicate standards were used for calibration (e.g., Amelia albite 
for Na, Si; microcline for K; Gates wollastonite for Ca; Alaska Anorthite for 
Al; synthetic fayalite for Fe; synthetic forsterite for Mg; synthetic TiO$_2$ 
for Ti; synthetic Mn-olivine for Mn; synthetic Cr$_2$O$_3$ for Cr).  
Microprobe analyses are given in Table~\ref{microprobe}, and the resulting 
compositions are given as chemical formulae in Table~\ref{tab:samples}.
Table~\ref{microprobe} also lists water contents, which
we determined  from near-IR spectra using the method in 
\citet{hofmeister09}.

\subsection{Viscosimetric Determination of Glass Transition Temperature}

As discussed in \S~\ref{silglass}, the glass transition temperature ($T_g$)
is essentially the temperature above which a given composition 
should form as or convert to crystalline solids, while solids formed 
below this temperature should be glassy.
We determine $T_{12}$ values as a proxy for $T_g$ 
for our silicate samples to provide an upper limit 
on temperature for models of glass formation in space.
$T_g$ depends on composition and cooling rate.  
%
%
Viscosity was measured over a range of temperatures using a Theta Instruments 
Rheotronic III parallel plate viscometer, following procedures described by 
\citet{whittington09}.
Viscosity is calculated from the measured longitudinal strain rate, known load 
and calculated instantaneous 
surface area, assuming perfect slip between sample and plates. The measurements are interpolated to find 
$T_{12}$ with an uncertainty of less than 2\,K.

\subsection{Spectroscopy}

Room temperature (18--19$^{\circ}$C) 
IR absorption spectra were acquired 
using an evacuated Bomem DA 3.02 Fourier transform 
spectrometer 
(FTIR)
at 1\,cm$^{-1}$ resolution 
The accuracy of the 
instrument is $\sim$0.01\,cm$^{-1}$. Far-IR data ($\nu < 650 {\rm cm}^{-1}$) 
were collected for five samples using a SiC globar source, a liquid helium cooled bolometer, 
and a coated mylar beam-splitter.
Mid-IR data ($\nu =$~450--4000\,cm$^{-1}$) were collected for all samples using a SiC globar 
source, a liquid nitrogen cooled HgCdTe detector, and a KBr beam-splitter.
The spectra were collected from powdered samples pressed in a diamond anvil 
cell (DAC).
The empty DAC was used as the reference spectrum which allows reflections to 
be removed.
%
The methodology is described by \citet{hofm03}.

Interference fringes exist in many of our spectra because the diamond faces are
parallel and are separated by a distance within the wavelength range studied.
The spacing is larger than film thickness. Fringes are associated with a
sideburst in the interferogram. Due to mathematical properties of Fourier
transforms, the fringes are convolved with the spectrum and therefore do not
affect peak shape or the parameters used to describe the peak (position, width
and height).

\section{Comparison of new laboratory spectra with previous laboratory and 
synthetic spectra}
\label{comparison}

\subsection{New laboratory spectra}
\label{newlabdata}

\begin{figure}[t]
\includegraphics[angle=270,scale=0.65]{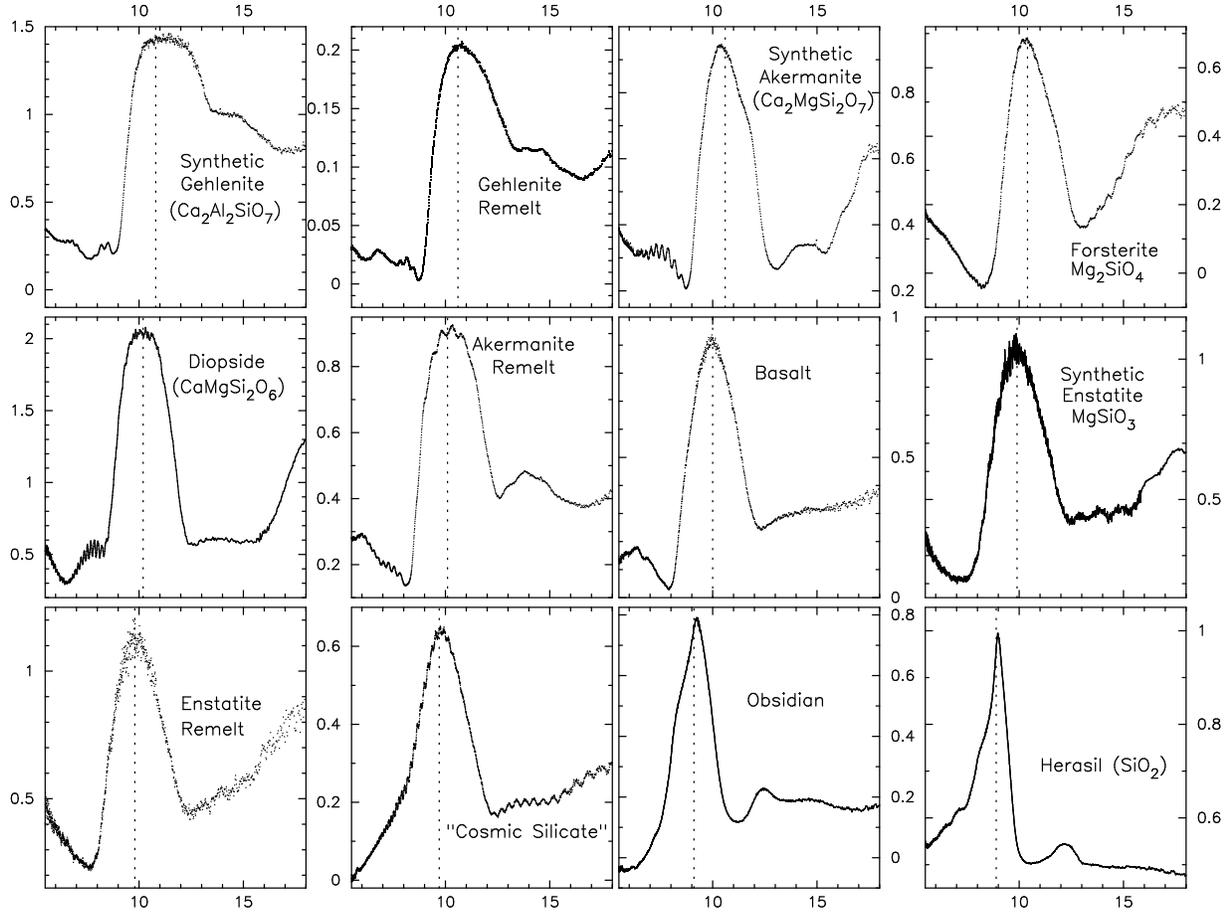}
\caption{\label{newlabdatafig} 
New laboratory absorbance spectra of glasses.
In all cases, 
$x$-axis is wavelength in $\mu$m; 
$y$-axis is absorbance. 
The dotted line indicates the barycentric position for each ``10\,$\mu$m'' 
feature.
The wavelength of these positions, along with the peak position and FWHMa are 
listed in Table~\ref{tab:samples3}.
The precise compositions for each sample are listed in 
Table~\ref{tab:samples}.}
\end{figure}

\begin{figure}[t]
\includegraphics[angle=270,scale=0.65]{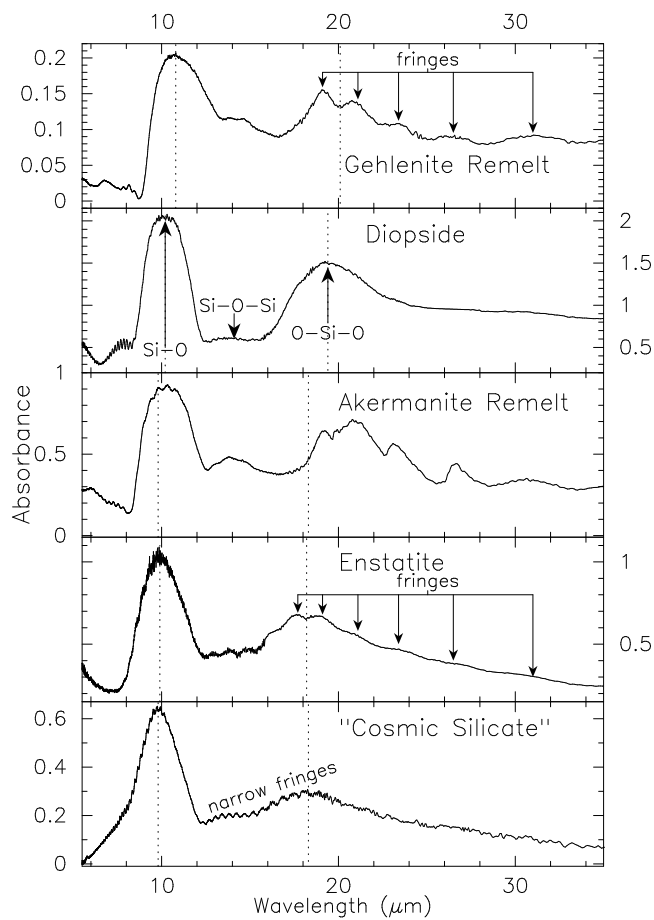}
\caption{\label{newlabdatafig20um} 
New laboratory absorbance spectra where data extends to far-IR.
In all cases, 
$x$-axis is wavelength in $\mu$m; 
$y$-axis is absorbance. 
The dotted line indicates the barycentric position for the ``10\,$\mu$m'' and ``18\,$\mu$m'' features.
The wavelength of these positions, along with the peak position and FWHMa are 
listed in Table~\ref{tab:samples3}.
The precise compositions for each sample are listed in Table~\ref{tab:samples}.
Interference wide fringes are indicated by arrows for gehlenite remelt and enstatite, 
while narrow fringes are marked on the spectrum of Cosmic Silicate.
The attributions for the observed features at $\sim10, 14,$ and $18\mu$m are indicated in the lower right panel.}
\end{figure}

The new laboratory spectra, shown in
Figure~\ref{newlabdatafig}, are the highest resolution spectra of 
silicate glasses to date. These data are available online from  
http://galena.wustl.edu/$\sim$dustspec/idals.html
%
The main observable parameters of astronomical spectra are the peak position 
of the absorption/emission features at $\sim 10\mu$m and $\sim 18\mu$m; their 
FWHMa and the ratio of their strengths (see e.g., \citeauthor{ohm92} and references therein). 
For instance, the ratio of band strengths between the 10 and 18\,$\mu$m 
features
in observations varies markedly, as do the peak positions. 
Therefore we have extracted the most important spectral parameters from our 
data using the NOAO onedspec package within the Image Reduction and Analysis 
Facility (IRAF).
The spectral parameters (the peak position, barycentric position, 
full width half maximum, FWHM, and equivalent width)  were determined for 
the $\sim10$ and $\sim18\mu$m  features. 
Figure~\ref{newlabdatafig} shows the barycentric position of the $\sim10\mu$m
feature for all samples, while 
Figure~\ref{newlabdatafig20um} shows the barycentric positions of both the 
$\sim 10$ and $\sim18\mu$m features for the five compositions for which far-IR 
data were collected.
The barycentric positions, along with the peak positions and FWHMa  from
absorbance measurements
are listed in Table~\ref{tab:samples3}.

There are multiple terms and symbols for the various ways in which absorption 
of light by solids is described. In order to prevent confusion, we will explain
precisely how each term we use is defined. This is particularly important 
for applying laboratory spectra to astrophysical studies because our spectra
are initially in the form of absorbance ($a$), but we typically use either
absorption efficiency (Q-values), optical depth or extinction in astronomy.

Transmittance, $T$ \citep[which is also referred to as 
{\em transmissivity} in][]{fox02} is defined as the ratio of the 
intensity of transmitted ($I_{\rm trans}$) and 
incident ($I_0$) light, i.e. $I_{\rm trans}/I_0$. Similarly reflectivity, $R$ 
and absorptivity, $A$ are the ratios of absorbed to incident light and 
reflected to incident light, respectively\footnote{note that 
{\em absorbance} is the reciprocal of the log of the {\em transmittance}, in
contrast to the absorptivity defined here; see also \citet{STH05}}. 

\[ I_0 = I_{\rm abs} + I_{\rm trans} + I_{\rm refl} \]

\[  \frac{I_{\rm trans}}{I_0} = 1 - \frac{I_{\rm abs}}{I_0} -  \frac{I_{\rm refl}}{I_0} \]

\[ T = 1 - R - A \]

\noindent
For simplicity we will assume the reflectivity is negligible or has been 
accounted for  
\citep[see][for how we can account for reflections]{hofmeister09} .
Then,

\begin{equation}
T = 1 - A
\end{equation}

\noindent
When light passes through a solid the absorption can be expressed as:

\begin{equation}
\label{tau1}
 I_x = I_0 e^{-\alpha L} 
\end{equation}

\noindent
where $L$ is pathlength or thickness of the solid sample and $\alpha$ is 
usually called the absorption coefficient, but is sometimes called 
opacity. In addition $\alpha = \kappa \rho$, where 
$\kappa$ is the mass absorption coefficient, and 
$\rho$ is the mass density.

\begin{equation}
\label{tau2}
T = e^{-\alpha L}= e^{-a} 
\end{equation}

\noindent
Absorbance, $a$, is the exponent in the decay of light due to absorption: 
$a = \alpha L = \kappa \rho L$.
Optical depth $\tau_\lambda$ of an absorbing material is defined by:

\begin{equation}
\label{tau3}
I_x = I_0 e^{-\tau_\lambda} 
\end{equation}

\noindent
\citep[From, e.g.][]{glass99}.

From equations~\ref{tau1}, \ref{tau2} and \ref{tau3} 
we see that the absorbance $a$ is similar to optical depth $\tau_\lambda$. 
In order to compare astronomical data in which we have a 
wavelength-dependent optical depth we can use absorbance 
(which is how we present our transmission spectra in Fig.~\ref{newlabdatafig}
and Fig.~\ref{newlabdatafig20um}
after accounting for surface reflections.)
Now we can relate the absorbance ($a$) and 
absorptivity ($A$) via equation~\ref{tau2}:

\[ A = 1 - e^{-a} \]

\noindent
To compare with some astronomical observations we still need to extract a 
version of the laboratory data that is comparable to the absorption efficiency,
$Q$-values. 
To get this we need to consider how $Q$-values are defined.
For a non-blackbody dust grain we define an absorption cross-section 
$C_{\rm abs}$ as the effective geometrical cross-section of the particle once 
we account for it not being a blackbody:

\[C_{\rm abs} = Q_{\rm abs} \times \Upsilon\]

\noindent
where $\Upsilon$ is the  geometrical cross-sectional area of a dust grain.
Now if we consider how the absorption cross section gives rise to absorption 
we get:

\begin{equation}
\label{sigma1}
\frac{I_{\rm abs}}{I_0} = A = C_{\rm abs} n L
\end{equation}

\noindent
where $n$ is the number density of absorbing particles and $L$ is the 
pathlength or thickness of the absorbing zone.

\begin{equation}
\label{rho}
 n = \frac{\rho}{M_{\rm mol} \times m_H} 
\end{equation}

\noindent
where $M_{\rm mol}$ is the molar mass of the solid and $ m_H$ is the mass of a 
hydrogen atom. Combining equations~\ref{sigma1} and \ref{rho} we get:

\begin{equation}
\label{Qeq}
Q_{\rm abs} = \frac{A \times M_{\rm mol} \times m_H}{ \Upsilon \rho L}
\end{equation}

\noindent
It is clear from equation~\ref{Qeq} that Q-values $\propto$ absorptivity such 
that:

\[ Q_{\rm abs} = \zeta A \]

\noindent
where, 

\[ \zeta =  \frac{M_{\rm mol} \times m_H}{ \Upsilon \rho L} \]

\begin{figure}[t]
\includegraphics[angle=270,scale=0.9]{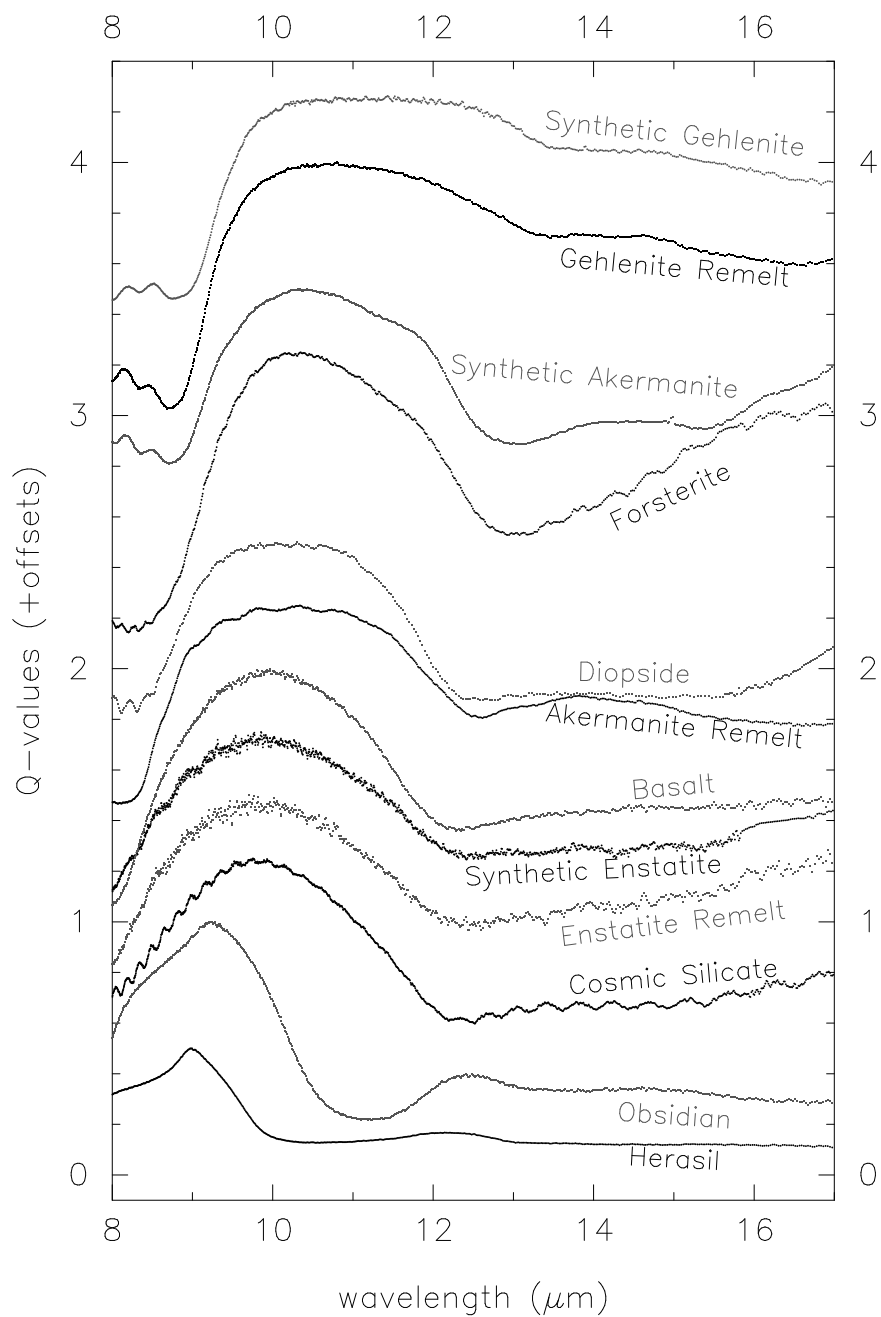}
\caption{\label{newlabdatafig2} 
Calculated Q values (absorption efficiency factors) for our samples
In all cases, 
$x$-axis is wavelength in $\mu$m; 
$y$-axis is $Q_{\rm abs}$. 
The spectra are plotted such that the barycentric position of the 
feature is reddest at the top and bluest at the bottom and are offset for 
clarity. The values for the offsets in $y$, along with the barycentric 
positions, peak position and FWHMa are 
listed in Table~\ref{tab:samples3}.}
\end{figure}

\noindent
Consequently, while the shape, peak position and FWHM of spectral features 
shown in $Q$-values will be identical to those for $A$, the absolute values 
depend on the pathlength and on the cross-section areas of a given grain 
distribution. The uncertainty in the thickness of our samples and in 
potential grain
size makes the absolute $Q$-values uncertain. Consequently we normalize our 
$Q$-value spectra to peak at unity.
%
%
%
These Q-value spectra  are stacked in Fig~\ref{newlabdatafig2} to demonstrate
the shift in both barycentric position and FWHM with composition.
Table~\ref{tab:samples3} also includes the FWHM in Q-value.



\subsection{How to compare disparate data sets}
\label{how2comp}

Here we compare the new laboratory spectra with those previously published 
to distinguish which factors are most important in determining spectral 
feature parameters. Previously published data are available as complex 
refractive indices ($n$ and $k$).
To make a fair comparison between the many 
available datasets we converted our absorbance data to the 
wavelength-dependent imaginary part of the complex refractive index ($k$) 
for each sample. This absorption index $k$ is chosen as the best 
comparison of different datasets, as it does not depend on grain sizes or 
shapes and can be extracted equally well from transmission or reflectance data 
\citep[see e.g.,][]{hofmeister09,fox02}.
Comparing absorption efficiency $Q_{\rm abs}$ requires 
assumptions about grain shape, which have been shown to affect the shapes and 
positions of spectral features \citep[e.g.][]{Min03,DePew06}. 
The conversion of the absorbance 
spectrum to $k$-values uses:

\[ k  = 2.303 a /(d 4 \pi \nu) \]

\noindent
where $a$ is the absorbance, $d$ is the sample thickness (in cm)
and $\nu$ is the 
frequency (wavenumber) in cm$^{-1}$. There is some uncertainty in the 
measurement of the sample thickness which is estimated to be 
0.5 -- 1.5\,$\mu$m thick \citep{HB06}. 
Consequently, for comparison our data has been 
normalized to peak at a $k$-value of 1.
In Figure~\ref{compare1} we compare spectra of
forsterite, enstatite, gehlenite and \r{a}kermanite composition 
``amorphous'' silicates.

\subsection{Comparison of new glass spectra with existing laboratory data.}
\label{comparesect}

\begin{figure}[t]
\includegraphics[angle=270,scale=0.63]{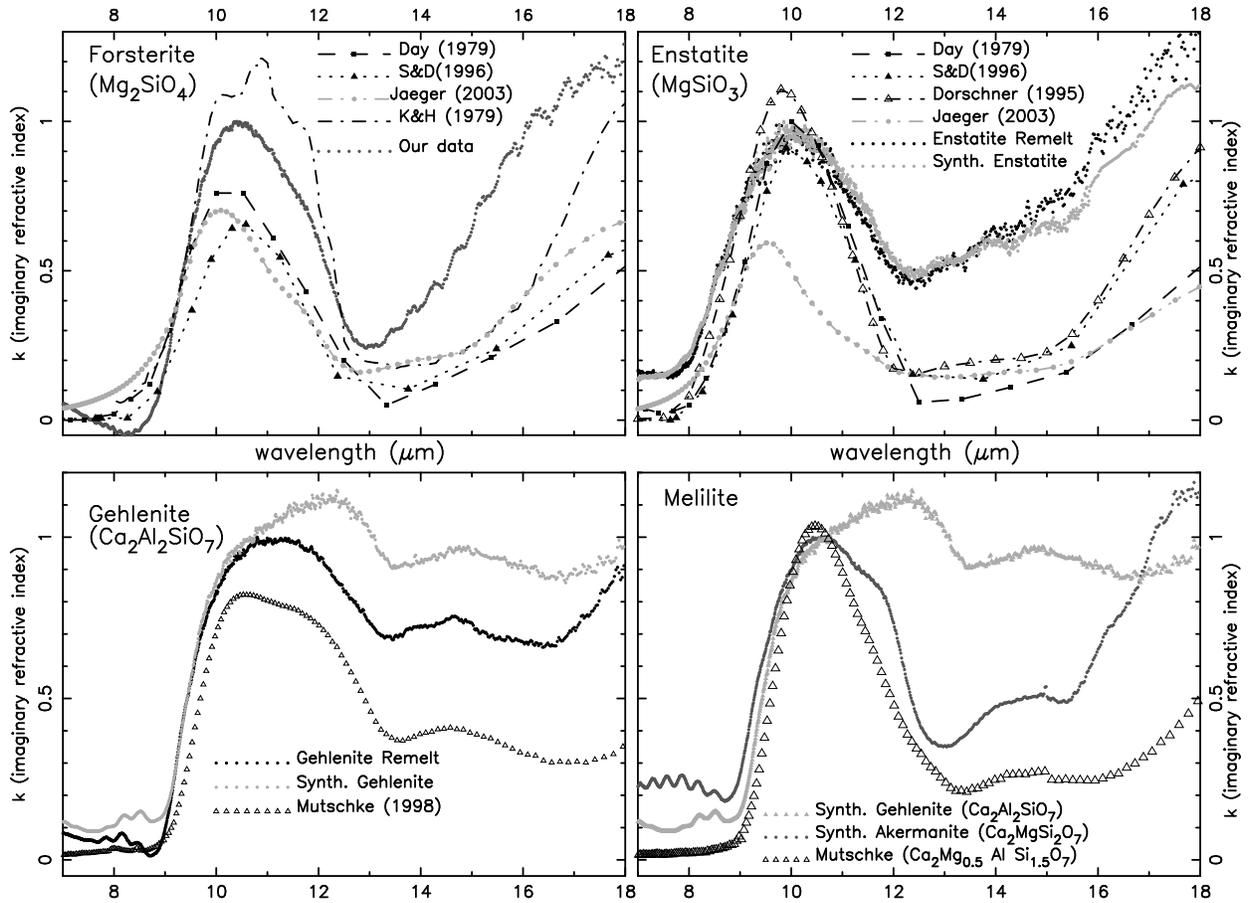}
\caption{\label{compare1} 
Laboratory optical data for amorphous silicates. 
Grey dots are our high-resolution spectra . 
S\&D96 = \citet{sd96}; 
Mutschke = \citet{mutschke98}; 
K\&H (1979) = \citet{kh79}. 
Symbols are the actual data points 
(except K\&H79, which was digitized from a figure). }
\end{figure}

Various synthesis techniques are associated with samples studied in the 
laboratory,
 as described in \S~\ref{prevlab} and listed in Table~\ref{litdata}.
%
Considering, for example, forsterite composition samples  
shown in the upper left panel of Figure~\ref{compare1},
it is clear that the spectral parameters vary even for ostensibly the same 
composition.
%
The ion irradiation technique used by  \citet{kh79} apparently
does not produce a fully amorphized sample, as this spectrum is closer to that 
of crystalline forsterite, which was their starting material.
The sample from \citet{day79} was produced via chemical vapor deposition; 
the \citet{sd96} sample was produced by laser ablation of crystalline samples;
and the sol-gel method was used by \citet{jager03}.

The spectra of three different samples produced by
chemical vapor deposition, laser ablation and sol-gel techniques  
are similar.
 However, although these samples may be ``amorphous'' they are not 
necessarily glassy. 
The samples generated by these three techniques may 
represent  chaotic silicates rather than the glassy silicates investigated herein.
%
The difference between previous samples and those presented 
here is most likely a combination of  density and porosity. 
Glasses should be less porous and denser than chaotic silicates.
%

Among the enstatite composition samples, those from \citet{day79} and 
\citet{sd96} are, again, similar to each other.
Samples produced by melting and quenching 
\citep[e.g. ours,][]{dorschner95} show some
variability in the spectral features. However, 
the peak positions and FWHMa are similar 
(see Tables~\ref{litdata} and \ref{tab:samples}). 
The difference in breadth may result from different 
cooling rates and fictive temperatures.
%
%
\citet{jager03} investigated amorphous MgO-SiO$_2$ solids prepared using the 
sol-gel method. The spectra vary markedly with changes in SiO$_2$ content, and 
hence with polymerization (Fig.~\ref{compare2}). 
However, these samples also contained a reported 0.8 to 1.2 wt.\% H$_2$O, 
which equates to $\sim$3 mol.\% H$_2$O, all dissolved as network-modifying 
hydroxyl (OH$^-$) ions (see \S~\ref{silglass}).
Glasses quenched from melts at atmospheric pressure 
\citep[e.g.\ those presented here and in][]{dorschner95}
contain much less water, typically 0.02--0.1wt.\%.
In previous laboratory studies it has been suggested that the changes in peak 
positions are due to differing silica contents, and hence polymerization 
states (NBO/T), i.e. forsterite has a redder peak position than enstatite, 
which in turn is redder than silica 
\citep[see e.g.,][and reference therein]{ohm92}. 
This is demonstrated in Fig.~\ref{compare2}.
Thus, even the modest levels of water remaining 
in samples prepared using the sol-gel
method will affect the structure of the silicate and thus its 
spectrum.
This makes precise interpretation of spectra from  sol-gel samples difficult, 
especially of the more 
silica-rich compositions whose structure will be the most affected by the 
incorporation water. 
Therefore the sample/material structure is not well known.
Spectral differences between the sol-gel MgSiO$_3$ of 
\citet{jager03} and the quenched MgSiO$_3$ glass of \citet{dorschner95} 
emphasize the point, especially around the 10$\mu$m feature 
(Fig.~\ref{compare1}).

The bottom left panel of Fig.~\ref{compare1} 
compares spectra for the gehlenite, the Al-rich endmember of the melilite series.
All data were produced by the melt-quench method. 
Gehlenite from this study closely matches spectra of the 
\citet{mutschke98} sample, while the sample we generated by melting a natural 
crystal of gehlenite (designated gehlenite remelt) shows significant 
differences which can be attributed to the deviation from 
stoichiometry
which results in large shifts in both Al/(Al$+$Si) and NBO/T
(see Table~\ref{tab:samples}).
The bottom right panel compares the mid-composition melilite (\r{A}k50Ge50) 
from 
\citet{mutschke98} with our end-member synthetic melilites. 
The peak position of the \citet{mutschke98} melilite is similar to our
Al-rich endmember, while the feature shape is closer to that of our Mg-rich endmember.

\citet{mutschke98} discussed the low contrast feature in the 12--16\,$\mu$m 
range in various aluminous silicates. 
This feature moves from $\sim14.5$\,$\mu$m in gehlenite towards shorter 
wavelengths, with the \r{a}kermanite feature peaking closer to 13.5\,$\mu$m
 (Fig.~\ref{compare1}).
\citet{mutschke98} did not investigate the \r{a}kermanite-rich melilites 
because their focus was on the effect of aluminium, but their findings, and 
those seen here may pertain to the carrier of the 
observed ``$\sim13$\,$\mu$m'' feature \citep[e.g.][and references therein]{sloan03}.

\begin{figure}[t]
\includegraphics[angle=270,scale=0.7]{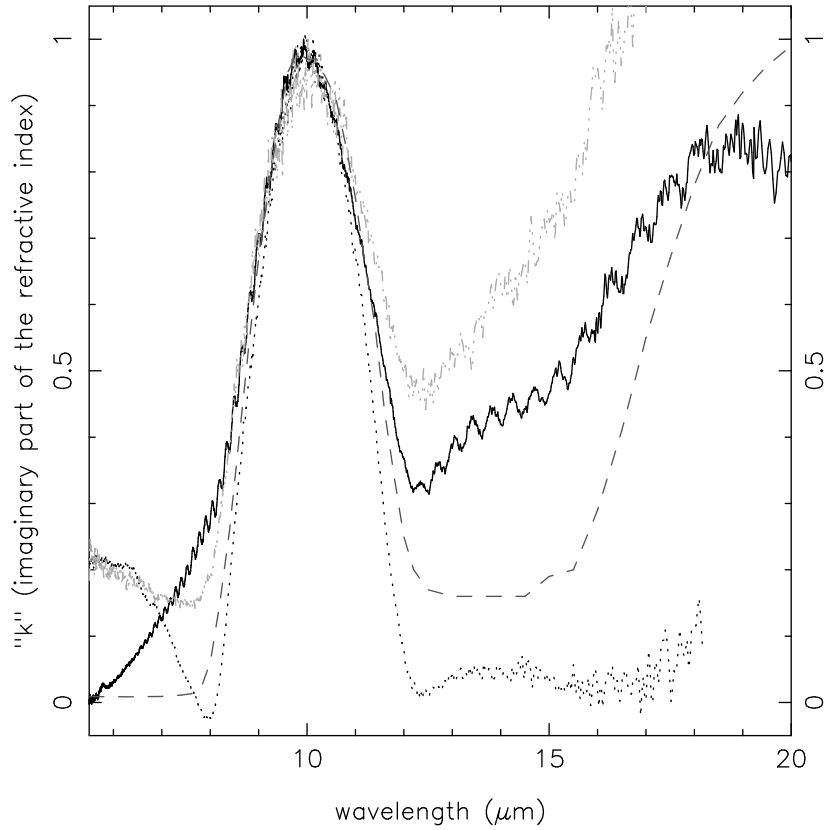}
\caption{\label{compare3} 
Comparison of ``dirty'' silicates.
Solid black line is our ``Cosmic'' silicate [Fe/(Mg+Fe)=0];
dotted black line is our ``Basalt'' [Fe/(Mg+Fe)=0.39];
dotted-dashed light grey line is Enstatite Remelt [Fe/(Mg+Fe)=0.07]
dashed dark grey line is the ``dirty'' silicate from  \citet{jager94}}
\end{figure}

Figure~\ref{compare3} shows a comparison between our ``cosmic silicate'', 
``basalt'' and the ``dirty silicate'' produced by \citet{jager94}. 
The 10\,$\mu$m features are very similar in peak position and FWHM, but the 
ratio of the $\sim$10 and $\sim18$\,$\mu$m features vary. This comparison 
shows that samples containing several elements give rise to very  similar 
10\,$\mu$m spectral features
even though they differ in Fe/(Mg+Fe), oxidation state, NBO/T and 
other compositional parameters.

In single crystal silicates, the Mg/Fe ratio affects the 
spectral features \citep[e.g.,][]{koike03,hp07}. \citet{dorschner95} 
investigated the effect of Mg/Fe ratio on the spectra of (Mg,Fe)SiO$_3$ 
glasses produced by quenching melts in air. The peak positions shift slightly, 
and the ratio of the 10\,$\mu$m and 18\,$\mu$m feature heights varies markedly 
(Fig.~\ref{compare2}). 
Substitution of Ca for Mg produces substantial broadening of the 10\,$\mu$m 
feature, consequently the peak (and barycenter) shift redward. 
The viscosity and glass transition data in Table~\ref{tab:samples}  
show that small 
impurities can have important effects on melt properties (and structure). For 
example, compare synthetic and remelted \r{a}kermanite and gehlenites. 
This is why it is important to consider Ca in pyroxenes, in addition to the 
En-Fs series.
Indeed based on the spectra presented here, the effect of calcium substitution 
is larger than that of iron.
%

\citet{dorschner95} reported that the 
FeO/Fe$_2$O$_3$ ratios of their samples were $\approx$1. 
Whether molar or weight ratio (unspecified in their paper), 
these samples do not have pyroxene stoichiometry, and thus have
a structure differing from that of our pyroxene glasses. Using their reported 
compositions, and assuming all Fe$^{3+}$ acts as a network former, 
calculated NBO/T values ranges from 2.1 for En95 to $\sim$0.8 for En50 and 
En40 glasses, considerably lower than the value of 2.0 for a true 
(Mg,Fe)SiO$_3$ composition. Furthermore, the effect of tetrahedral Fe$^{3+}$ 
on neighboring Si--O bonds, which give rise to the 10\,$\mu$m feature, may be 
sufficient to produce marked changes in the shape and peak position of the 
feature even if Mg-Fe substitution does not. Therefore existing lab data on 
glasses do not allow astronomical spectra to be interpreted reliably in terms 
of either M$^{2+}$/Si or Mg/Fe ratio, which are important tools for discriminating 
between competing dust formation models. 
We will address the roles of oxidation state and Al-content in silicates in 
future papers.

\subsection{Comparison with Synthetic Spectra}

As discussed in \S~\ref{prevlab} the most popular silicate spectral data used 
in astronomy are the synthetic optical functions (complex refractive 
indices/complex dielectric functions) from \citeauthor{dl84} and \citeauthor{ohm92}.
In the spectral regions considered here, both groups derived
dielectric functions from astronomical observations.
\citeauthor{dl84} used observations  of the ISM
while 
\citeauthor{ohm92} produced two sets of optical constants 
designated as warm O-deficient and cold O-rich. The warm O-deficient is 
intended to match circumstellar dust features where the \citet{NuthHecht} 
predicted that non-stoichiometric silicates form in the outflows. 
The cold O-rich silicate is meant to represent dust formed in molecular clouds, 
where the cool temperatures and slow dust formation lead to stoichiometric 
compositions. 
For the circumstellar (O-deficient warm), 
opacities are derived by \cite{vk88} based on averaging approximately
500 IRAS LRS spectra of evolved stars. However, this approach is less than 
ideal. It is well known that the observed ``silicate'' features vary markedly 
from source to source \citep[see e.g.][]{NuthHecht,mutschke98,speck00} and thus 
averaging 
can smear out such differences and give rise to an opacity that is close to 
matching many objects and actually matches none.

To provide a fair comparison, parameters for these synthetic 
silicate spectra have been included in Table~\ref{litdata}.
Whereas the \citeauthor{dl84} spectra provide a reasonable match to the 
features observed in the ISM, the spectra from \citeauthor{ohm92} provides a 
narrower 
$\sim10\mu$m feature which matches the circumstellar silicate features more 
closely \citep[e.g.][and references therein]{Sargent2010}. However, neither 
\citeauthor{dl84} nor \citeauthor{ohm92} provide perfect matches to the observed 
astronomical mid-IR features; nor do they reflect the diversity of features 
seen in astronomical environments. 
%
Although synthetic spectra do not match astronomical data perfectly, 
they are widely used and come close to matching many astronomical observations.
Consequently, comparing our new laboratory data to these synthetic spectra allows us to assess
whether any true glasses show promise as carriers of observed silicate features.

\begin{figure}[t]
\includegraphics[angle=270,scale=0.75]{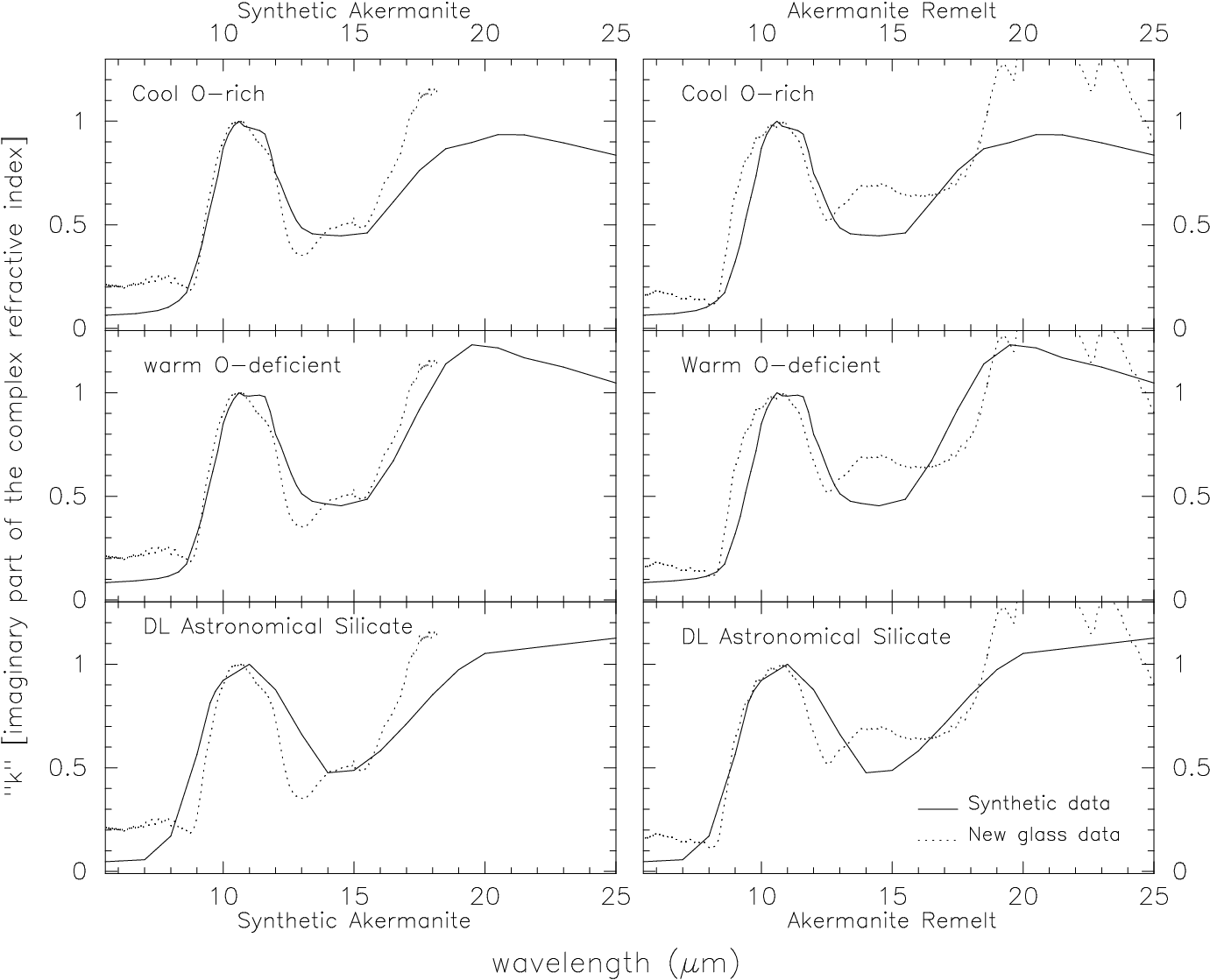}
\caption{\label{synthcomp1} 
Comparison of synthetic spectra from \citeauthor{dl84} and \citeauthor{ohm92}
with new laboratory spectra on \r{a}kermanite.
$x$-axis is wavelength in $\mu$m; $y$-axis is the imaginary part of the 
complex index of refraction ($k$).
{\em Left panels}: synthetic \r{a}kermanite;
{\em right panels}: \r{a}kermanite remelt;
{\em top row:} comparison to cool oxygen-rich silicate from OHM;
{\em middle row}: comparison to warm oxygen-deficient silicate from OHM;
{\em bottom row}: comparison to DL astronomical silicate.
Solid lines are the synthetic spectra; dotted lines are the new laboratory 
spectra. 
}
\end{figure}

\begin{figure}[t]
\includegraphics[angle=270,scale=0.75]{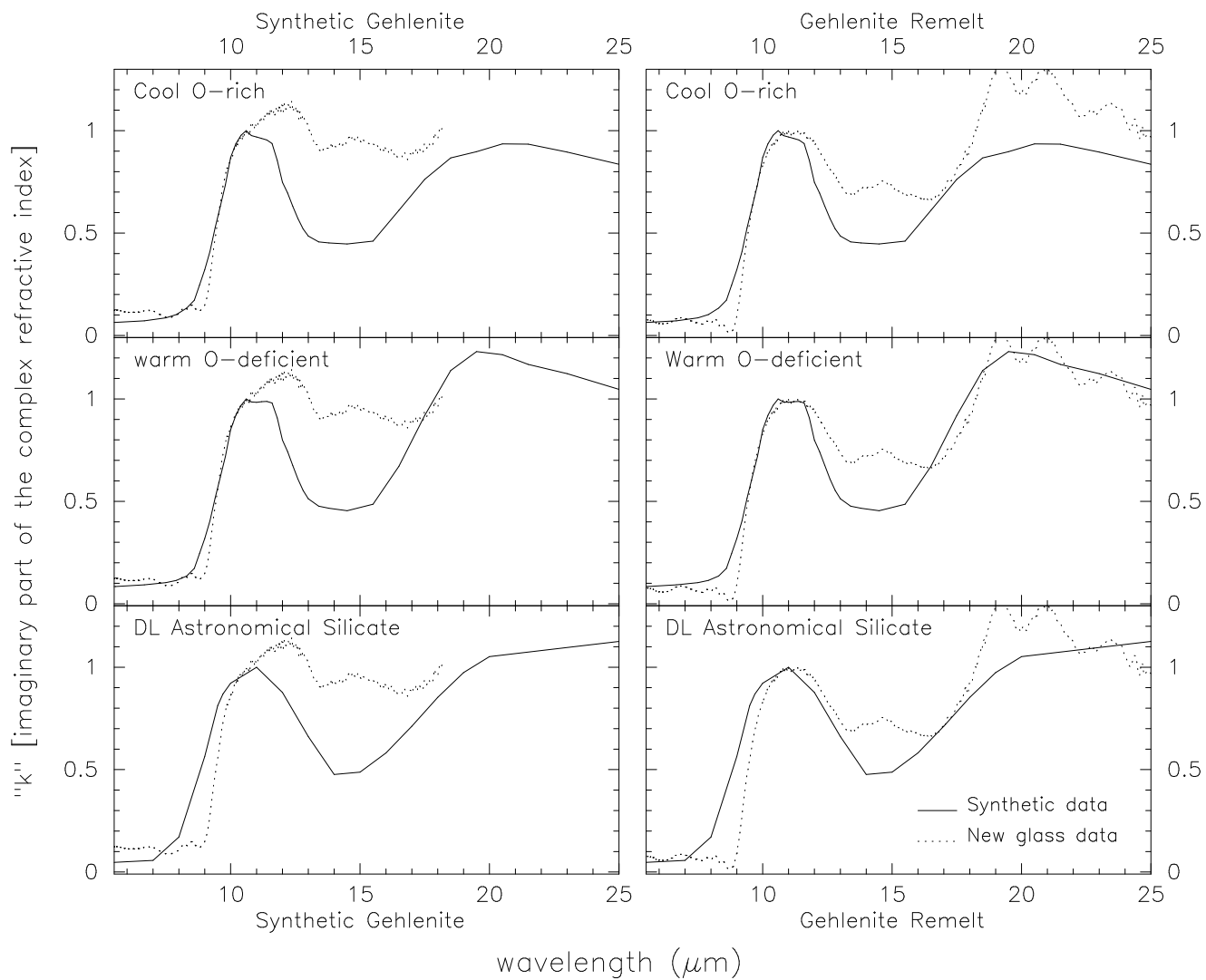}
\caption{\label{synthcomp2}
Comparison of synthetic spectra from \citeauthor{dl84} and \citeauthor{ohm92}
with new laboratory spectra on gehlenite.
$x$-axis is wavelength in $\mu$m; $y$-axis is the imaginary part of the 
complex index of refraction ($k$).
{\em Left panels}: synthetic gehlenite;
{\em right panels}: gehlenite remelt;
{\em top row:} comparison to cool oxygen-rich silicate from OHM;
{\em middle row}: comparison to warm oxygen-deficient silicate from OHM;
{\em bottom row}: comparison to DL astronomical silicate.}
\end{figure}

\begin{figure}[t]
\includegraphics[angle=270,scale=0.75]{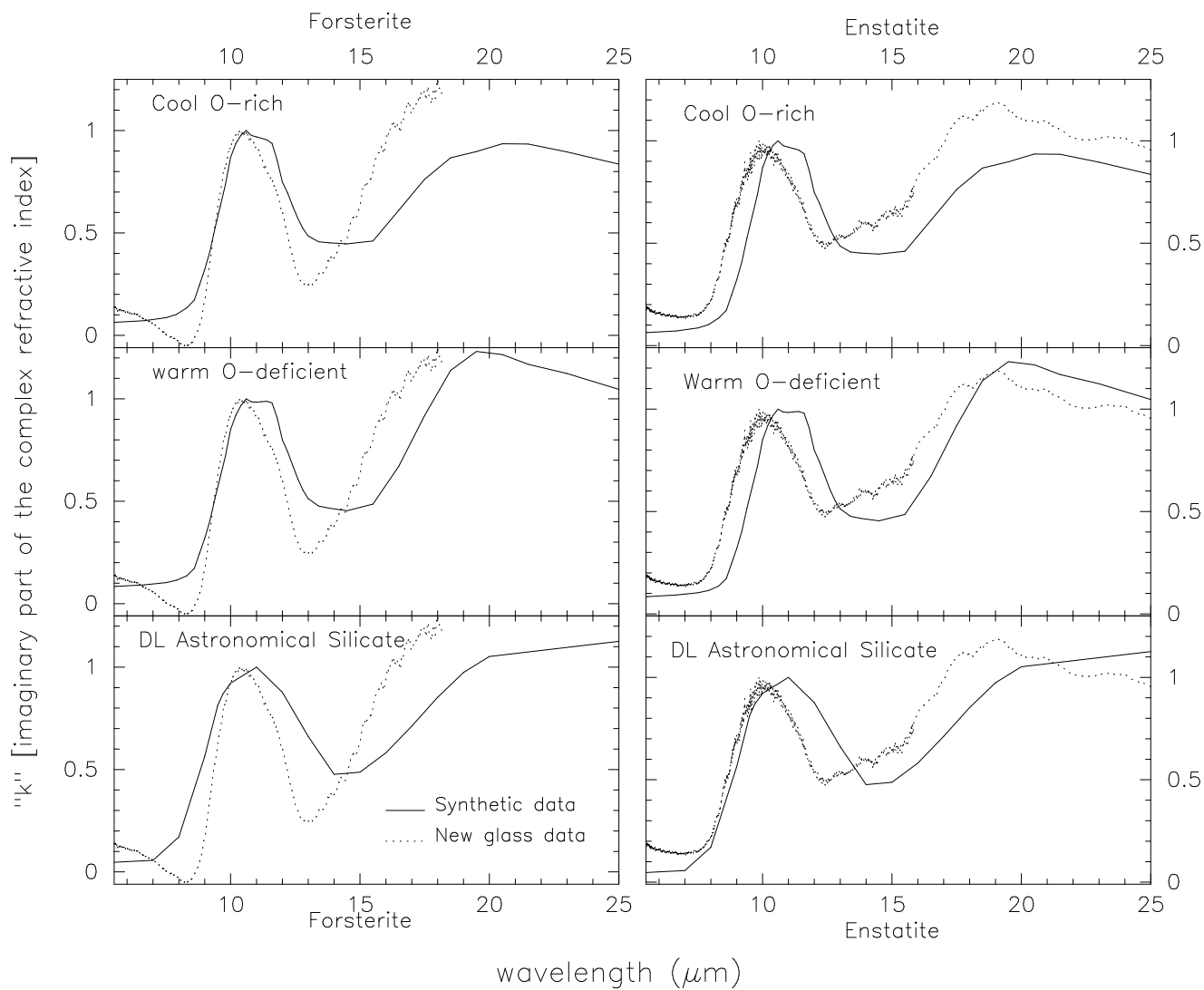}
\caption{\label{synthcomp5}
Comparison of synthetic spectra from \citeauthor{dl84} and \citeauthor{ohm92}
with new laboratory spectra on forsterite and enstatite.
$x$-axis is wavelength in $\mu$m; $y$-axis is the imaginary part of the 
complex index of refraction ($k$).
{\em Left panels}: forsterite;
{\em right panels}: enstatite;
{\em top row:} comparison to cool oxygen-rich silicate from OHM;
{\em middle row}: comparison to warm oxygen-deficient silicate from OHM;
{\em bottom row}: comparison to DL astronomical silicate.}
\end{figure}

\begin{figure}[t]
\includegraphics[angle=270,scale=0.75]{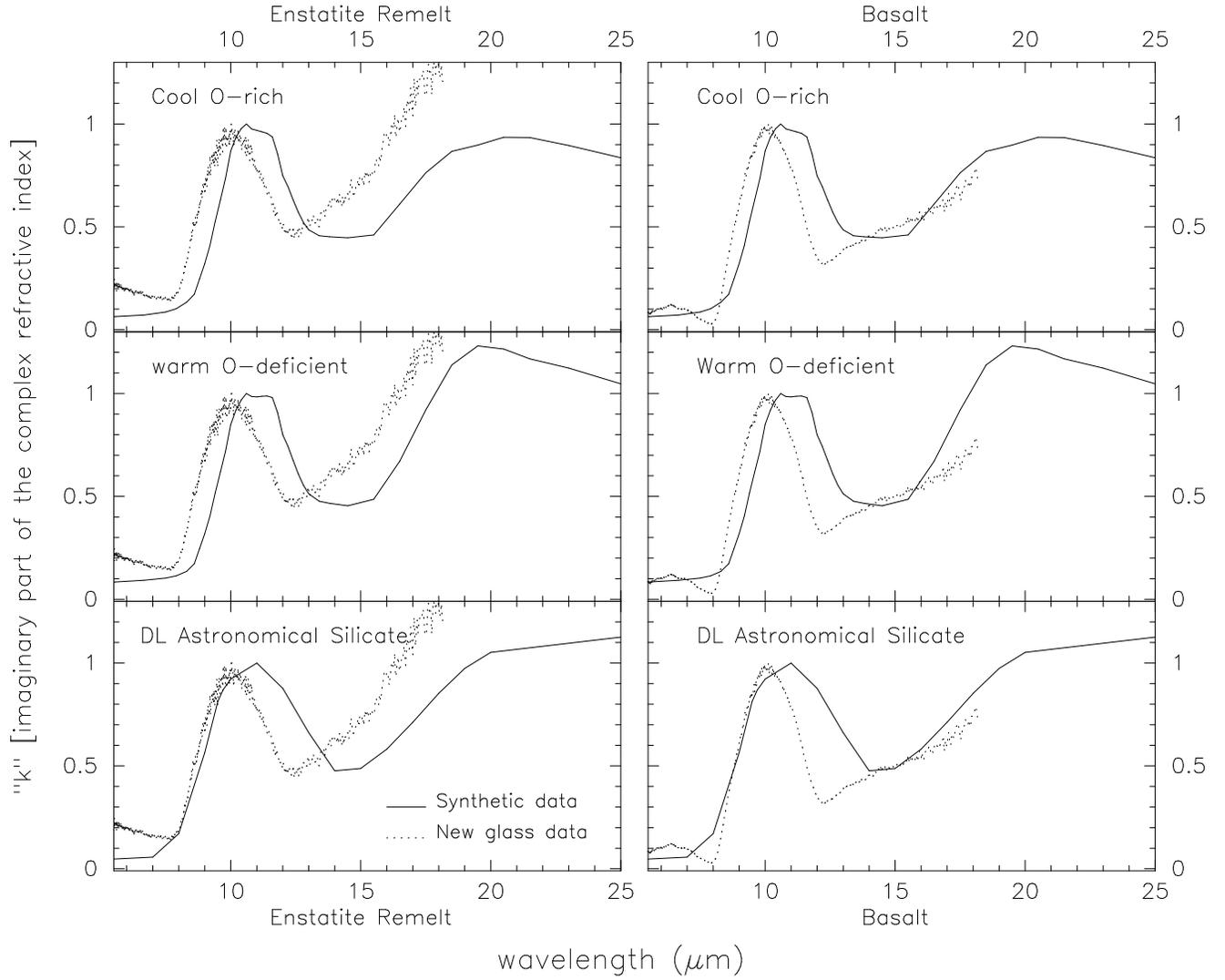}
\caption{\label{synthcomp3}
Comparison of synthetic spectra from \citeauthor{dl84} and \citeauthor{ohm92}
with new laboratory spectra with Enstatite Remelt and Basalt
$x$-axis is wavelength in $\mu$m; $y$-axis is the imaginary part of the 
complex index of refraction ($k$).
{\em Left panels}: enstatite remelt
{\em right panels}: basalt;
{\em top row:} comparison to cool oxygen-rich silicate from OHM;
{\em middle row}: comparison to warm oxygen-deficient silicate from OHM;
{\em bottom row}: comparison to DL astronomical silicate.}
\end{figure}

\begin{figure}[t]
\includegraphics[angle=270,scale=0.75]{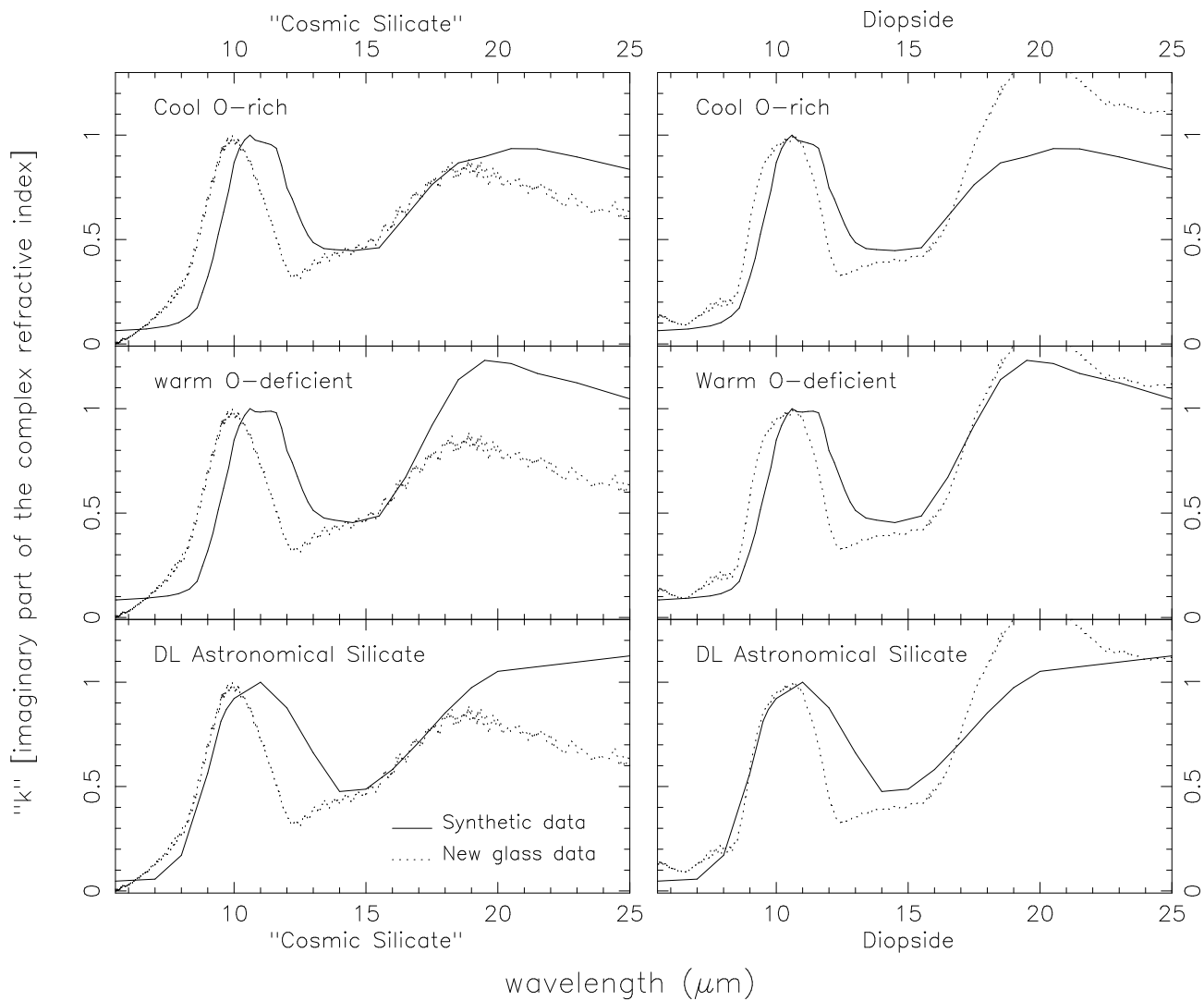}
\caption{\label{synthcomp4}
Comparison of synthetic spectra from \citeauthor{dl84} and \citeauthor{ohm92}
with new laboratory spectra on ``Cosmic Silicate'' and Diopside.
$x$-axis is wavelength in $\mu$m; $y$-axis is the imaginary part of the 
complex index of refraction ($k$).
{\em Left panels}: Cosmic silicate;
{\em right panels}: diopside;
{\em top row:} comparison to cool oxygen-rich silicate from OHM;
{\em middle row}: comparison to warm oxygen-deficient silicate from OHM;
{\em bottom row}: comparison to DL astronomical silicate.}
\end{figure}

Figures~\ref{synthcomp1} to \ref{synthcomp4} show how the spectral properties 
of the new laboratory samples compare to those of synthetic silicates from
\citeauthor{dl84} and \citeauthor{ohm92}. In all cases the comparison is made with the 
absorption index $k$, which is the imaginary part of the complex 
refractive index, and was chosen to avoid complications
arising from grain shape and grain effects (see \S~\ref{how2comp}).
None of the samples presented here precisely match the spectral features of the synthetic 
silicates. In particular, the 10$\mu$m feature in the spectra from \citeauthor{dl84} 
is too broad to be matched by any of the laboratory samples, with the possible 
exception of gehlenite (both ``synthetic'' and ``remelt''; see 
Fig.~\ref{synthcomp2}, bottom row). 
However, the relative cosmic abundances calcium and aluminium make this an 
unlikely attribution for the ISM dust.
It is usually assumed that astronomical 
silicates are largely comprised of  ``amorphous'' Mg-rich  olivine and pyroxene compositions.  
Therefore, we compare the interstellar feature as represented by \citeauthor{dl84} with 
such compositions in Fig.~\ref{synthcomp5}--~\ref{synthcomp3}. 
The spectra of the four Mg-rich pyroxene-like glasses (Synthetic Enstatite,
 Enstatite Remelt, Cosmic Silicate and Basalt) have grossly similar spectra.
In all four cases, the laboratory data match the blue side of the \citeauthor{dl84} feature, 
but are too narrow and fail to match the red side.
This is also true for the broader diopside feature (Fig.~\ref{synthcomp5}).
While forsterite composition glass has a redder feature than 
the pyroxenes, it is still too narrow and still not red enough to match 
the \citeauthor{dl84} 10$\mu$m feature.

If glasses do produce the astronomical 10$\mu$m band, the excess
breadth needs accounting for and may suggest processing of dust in the 
ISM \citep{NuthHecht}.
The physical changes to the dust that give rise to the broader interstellar 
feature remain unknown, but there are numerous interpretations.
In circumstellar environments the red-side broadening of the 10\,$\mu$m 
feature has been interpreted as being due to
oxide inclusions \citep[e.g.,][]{speck00}; 
increasing crystallinity \citep[e.g.,][]{sylvester98,bouwman01}; 
changes in grain shape \citep[e.g.,][]{min07}; or 
grain porosity \citep[e.g.,][]{henning93,vh08}.
The silicate feature is fairly constant and broad for the diffuse ISM, but varies more and is 
narrower for molecular clouds 
\citep[see e.g.][and references therein]{vanbreeman}. These variations 
have been attributed to a combination of grain agglomeration and 
ice mantle formation \citep[see][]{chiar06,chiar07,mcclure08,vanbreeman}, but 
the results of those studies depend on the optical properties input into 
their models.

The  spectra of \citeauthor{ohm92} have been used somewhat successfully in 
modeling circumstellar silicate features 
\citep[e.g.][and references therein]{Sargent2010}.
Of the new laboratory glasses, only forsterite comes close to matching the \citeauthor{ohm92} synthetic spectra  (Fig.~\ref{synthcomp4}), 
but there is still a problem with the breadth and redness of the feature.
Most other compositions
have 10$\mu$m features that are too blue. However, 10\,$\mu$m  silicate emission features seen in AGB stars spectra 
 and observed by ISO and IRAS are consistently slightly bluer than those in the synthetic spectra
\citep{wheeler07,wheeler09}. 
Consequently our samples may provide better matches than the OHM synthetic spectra.


\section{Discussion}

\subsection{Amorphousness at different scales: Glasses vs. Nanocrystalline solids}

\begin{figure}[t]
\includegraphics[angle=0,scale=0.6]{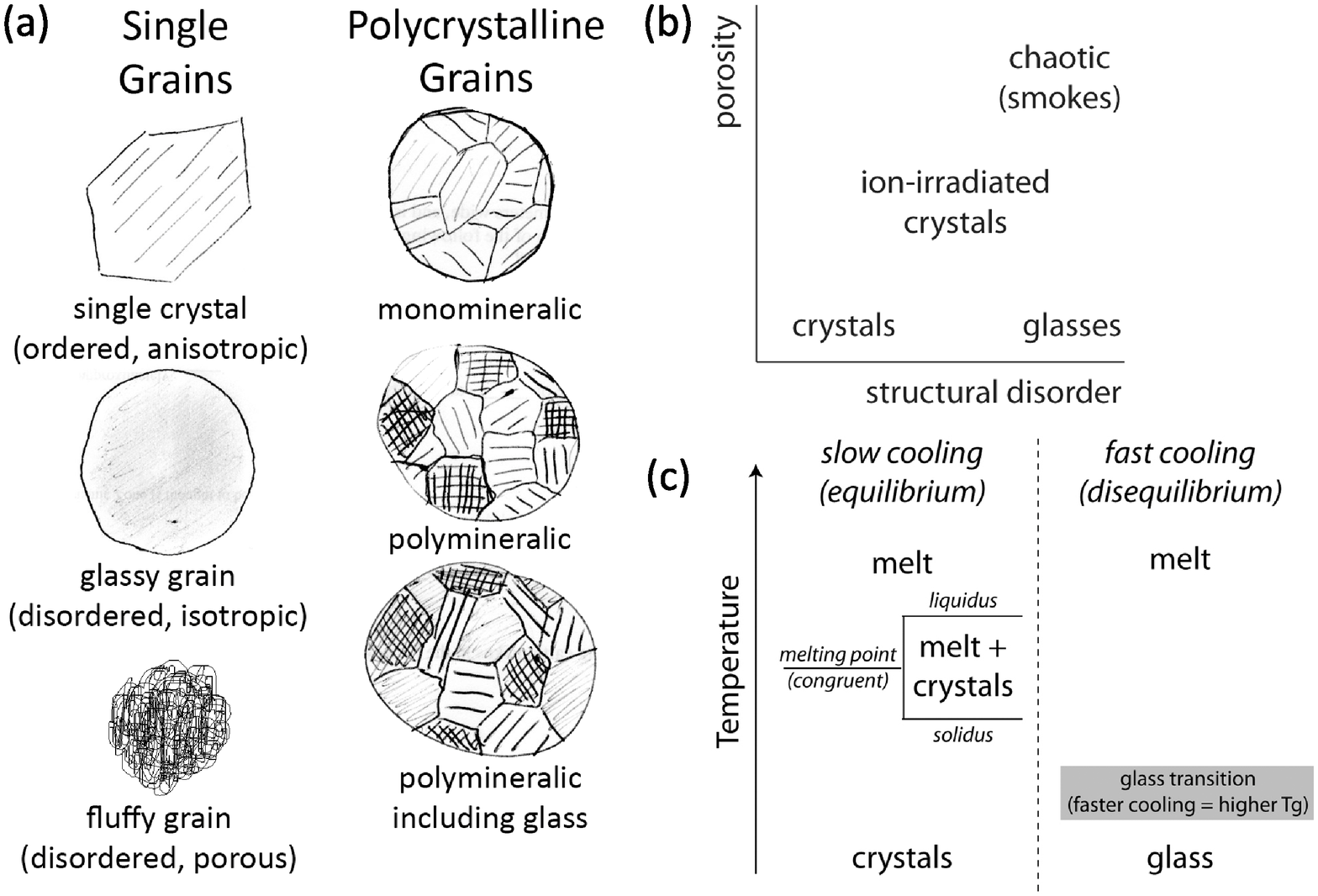}
\caption{\label{fig:XtalC} 
(a) Cartoon of possible grain structures, including single and polycrystalline 
grains. 
(b) Schematic of grain porosity vs structural disorder for different 
experimental sample materials. 
(c) Simple phase diagram illustrating the location of the glass transition 
range, below the melting point of crystalline materials. Which side of the 
phase diagram is relevant depends on the cooling rate of the system.  }
\end{figure}

We have shown that differences exist between spectra of amorphous samples of 
ostensibly the same composition  due to structural differences 
arising from synthesis methods.
In this section we discuss the range of structures that  
may be considered amorphous, and other factors such as composite or 
polycrystalline grains that may produce different spectral features 
to those obtained from studies of single crystals.
The range of structures is shown schematically in Fig.~\ref{fig:XtalC}.

Disordered (but not glassy) silicates may form at temperatures above $T_g$ by
ion bombardment of initially crystalline materials. 
Depending on the extent of damage done by ion radiation, an initially 
crystalline sample may be amorphized and could be indistinguishable 
spectroscopically from a truly glassy sample \citep[see e.g.][]{Demyk2004}.
As discussed in \S~\ref{prevlab}, limited heavy ion irradiation may not 
completely amorphize a crystalline sample \citep[e.g.][]{kh79}. 
Consequently ion-irradiation of initially crystalline material should 
lead to a continuum of structures ranging from perfect crystals to 
completely disordered, with the accompanying range of spectral features. 
Furthermore \citet{Demyk2004} showed that ion irradiation leads to more 
porous samples 
than simple splat-quenched glasses. 
While lower densities should not directly affect the spectra,
voids give rise to extra reflections which increases extinction and 
feature widths 
through increased scattering even though absorption is unchanged.
The structures of ion-irradiated crystals may be more similar to smokes than 
glasses because of ion damage (see Fig.~\ref{fig:XtalC}). 

Another potential carrier of the observed silicate spectral features is 
polycrystalline silicate. 
Spectra of single composition (monomineralic) polycrystals will be 
affected by multiple scattering which should broaden and smear out the 
features as demonstrated for hematite in \citet[][]{icarus}. 
%
%
%
During annealing and  crystallization of an initially glassy grain  
the final structure 
would depend strongly on composition. For a glass of pure enstatite 
composition, one may expect either a single crystal of enstatite or a 
mono-mineralic polycrystalline agglommeration (if crystallization starts at 
more than one point in the grain.) However, since glassy grains are 
expected to form to include all the atoms in the outflowing gas, fractional 
crystallization is possible. For instance, our ``cosmic'' silicate has a bulk 
composition close to enstatite, but contains a significant amount of other 
elements. Consequently we might expect an annealed sample of this composition 
to be poly-mineralic polycrystals with a large abundance of enstatite 
crystals.
Depending on the spectral contributions from other constituents of the 
polymineralic grain, we may not perceive sharp crystalline 
features in the spectrum. This is beyond the scope of the present work but 
will be investigated in the future.

Given that there are multiple mechanisms for the formation of both amorphous 
and crystalline silicate grains, and that processing can lead from one to the 
other and vice versa, it is possible that many types of grains represented on 
Fig.~\ref{fig:XtalC} are found in space.

\subsection{Potential application to astronomy}
\label{astroapp}


AGB stars present an interesting environment in which to study dust formation 
because they are relatively benign and the stability of CO molecules 
simplifies the chemistry. Whereas many AGB stars exhibit the $\sim10\mu$m 
feature, this feature varies in peak position and shape from object to object,
 and even temporally within a single object \citep{speck00,sloan03,monnier98}.
\citet{speck00} showed that several Galactic objects show 
silicate features peaking as short as 9.2\,$\mu$m, while
\citet{swt08} found a very red 10\,$\mu$m absorption feature in the spectrum 
of an obscured AGB star. While the synthetic spectra are commonly used to 
model the observed features, they rarely match the details of peak position 
and width. The new laboratory data presented here provide a framework to 
interpret the observed variations in AGB star spectral features. For example, 
the red feature seen by \citet{swt08} can be best matched by something 
forsteritic, while the observed 9.2$\mu$m feature needs  very silica-rich 
dust.


Recent studies by \citet{chiar06,chiar07}, \citet{mcclure08} and 
\citet{vanbreeman} have shown that the shape and peak position of the classic 
10\,$\mu$m interstellar silicate absorption feature
 varies depending on the line of sight. Whereas the $\sim10\mu$m feature 
remains the same for all diffuse lines of sight, its shape and position varies 
once the line of sight includes a molecular cloud. This has been attributed to 
a combination of grain agglomeration and ice mantle formation within the 
molecular clouds ({\em op.\ Cit.}).
%
The diffuse ISM is well characterized by 
\citeauthor{dl84} but for molecular clouds this $\sim$10\,$\mu$m feature is 
too broad \citep{vanbreeman}.
The competing hypotheses explaining the molecular cloud spectra could be 
tested by comparing  with the glass spectra presented herein.



The importance of dust to astrophysical processes cannot be over stated. 
For example, 
observations of high redshift ($Z>7$) galaxies and quasars demonstrate that 
there was copious dust produced by the time the Universe was $\sim$700 million 
years old \citep[e.g.,][]{sugerman06,dwek07}. 
Furthermore, in Active Galactic Nuclei (AGN), the observed silicate absorption 
feature is shifted to peak at a longer wavelength, and is broader than that 
observed in our own Galaxy. This spectral shift has been attributed 
to calcium-aluminum rich silicates \citep{jaffe04} or porous particles 
\citep{li08}.
Moreover, data from Spitzer (SAGE-IRS Legacy program) show a number of both 
AGB and YSO 
sources with remarkably blue silicate features. Given that Mg, Fe and Si are 
formed in different nucleosynthetic processes, the abundance ratios of these 
elements do not necessarily scale with metallicity. 
Disentangling dust formation mechanisms through observations of dust requires 
the optical properties of a range of silicate samples of 
varying Mg- Fe- and Si- contents and other components as provided here. 
The work presented here is a subset of a larger study and further sample 
compositions and structures
will be presented in the near future.

\section{Conclusions}

We have presented new laboratory spectra of astrophysically relevant silicate 
glasses and compared them to existing data in the literature. 
We have shown that
(1) ``disordered'' is not synonymous with glassy. In addition to structural 
disorder, porosity also affects spectral features.
(2) Sample preparation and characterization are important.
(3) We confirm the general trend of decreasing peak wavelength with increasing 
polymerization for the $\sim$10\,$\mu$m feature. However the scatter about 
this overall trend indicate that other compositional factors must be important.
(4) Nothing quite matches the diffuse ISM in peak position and breadth. 


Spectral parameters of disordered silicates are sensitive to 
composition and sample synthesis techniques, which reflect degree of disorder,
porosity, 
oxidation states and water content. To understand dust formation we must 
disentangle these parameters through further systematic study of major 
compositional series using high resolution spectroscopy on thoroughly 
characterized samples.
Further studies of the parameters will follow. 
These new data can be used for interpretation of more esoteric 
environments e.g. high redshift galaxies and novae.


\acknowledgements
This work is supported by
NSF AST-0908302 (A.K.S. and A.G.W.) and NSF AST-0908309 (A.M.H.)
and by
NSF CAREER AST-0642991 (for A.K.S.) and
NSF CAREER EAR-0748411 (for A.G.W.).
We would like to thank Bryson Zullig and Josh Tartar for their help with this 
work.






\clearpage

\clearpage

\clearpage

\clearpage

\clearpage

\clearpage

\clearpage

\clearpage
\begin{figure}[t]
\includegraphics[angle=270,scale=0.7]{f8.ps}
\caption{\label{compare2} 
Laboratory spectra demonstrating changes in peak position and shape with 
composition. 
a) {\em top panel}: the effect of SiO$_2$ content, data from \citet{jager94} 
and \citet{dorschner95}; 
b) {\em middle panel}: the effect of SiO$_2$ content, our  
data on Mg-endmember glasses;
c) {\em bottom panel}: the effect of Fe content on ``pyroxene'' glasses, 
where En\# = 100 $\times$ Mg/(Mg+Fe) in the glass. 
The iron-bearing glasses are substantially oxidized. 
Data from \citet{dorschner95}. }
\end{figure}

\clearpage

\clearpage

\clearpage

\clearpage

\clearpage

\clearpage



\clearpage
\begin{table}
\caption{Laboratory spectral data from astronomical literature}
\footnotesize
\label{litdata}
\begin{tabular}{p{2.25cm}@{\hspace{2mm}}p{3cm}@{\hspace{2mm}}p{2cm}@{\hspace{2mm}}p{1.5cm}@{\hspace{2mm}}p{2cm}@{\hspace{2mm}}p{1.5cm}@{\hspace{2mm}}p{2cm}c}
\hline
\raggedright Citation & \raggedright Sample Preparation Technique & \centering{Sample Composition}  & \centering{Barycenter ($\mu$m)} & \centering{FWHM ($\mu$m)}	& \centering{Barycenter ($\mu$m)} & \centering{FWHM ($\mu$m)}& \\ 
\hline
\raggedright \citet{day79}  &  \raggedright chemical vapor condensation  &  \centering{  Fo100  }  &  \centering{  10.2  }  &  \centering{  1.9  }  &  \centering{  21.8  }  &  \centering{  11.2  }  &  \\  
\vspace{-5mm}
      
  &    &  \centering{  En100  }  &  \centering{  10.0  }  &  \centering{  1.8  }  &  \centering{  20.9  }  &  \centering{  6.9  }  &  \\        
  &            &        &        &        &        &  \\        
\raggedright \citet{sd96} &  \raggedright laser ablation of crystalline samples  &  \centering{  Fo100  }  &  \centering{  10.4  }  &  \centering{  1.8  }  &  \centering{  22.3  }  &  \centering{  17.7  }  &  \\        
  &    &  \centering{  En100  }  &  \centering{  10.0  }  &  \centering{  1.8  }  &  \centering{  22.5  }  &  \centering{  17.3  }  &  \\        
  &            &        &        &        &        &  \\        
\raggedright \citet{kh79}   &  \raggedright ion-irradiation of crystalline samples  &  \centering{  Fo100  }  &  \centering{  10.5  }  &  \centering{  2.2  }  &  \centering{  18.0  }  &  \centering{  2.5  }  &  \\        
  &            &        &        &        &        &  \\        
Jena  &  \raggedright Melting and Quenching  &  \centering{  Fo50  }  &  \centering{  10.3  }  &  \centering{  2.3  }  &  \centering{  20.5  }  &  \centering{  11.5  }  &  \\        
  &    &  \centering{  En50  }  &  \centering{  9.8  }  &  \centering{  2.6  }  &  \centering{  20.0  }  &  \centering{  7.3  }  &  \\        
\raggedright \citet{dorschner95}&  \raggedright Melting and Quenching  &  \centering{  En100  }  &  \centering{  9.8  }  &  \centering{  1.8  }  &  \centering{  20.4  }  &  \centering{  9.8  }  &  \\        
  &    &  \centering{  En80  }  &  \centering{  9.8  }  &  \centering{  2.3  }  &  \centering{  20.2  }  &  \centering{  9.1  }  &  \\        
  &    &  \centering{  En60  }  &  \centering{  9.9  }  &  \centering{  2.5  }  &  \centering{  20.3  }  &  \centering{  9.3  }  &  \\        
  &    &  \centering{  En40  }  &  \centering{  9.7  }  &  \centering{  2.7  }  &  \centering{  20.4  }  &  \centering{  8.2  }  &  \\        
\raggedright \citet{jager03}   &  \raggedright Sol-Gel  &  \centering{  Fo100  }  &  \centering{  10.0  }  &  \centering{  2.2  }  &  \centering{  17.7  }  &  \centering{  4.0  }  &  \\        
  &    &  \centering{  En100  }  &  \centering{  9.5  }  &  \centering{  2.0  }  &  \centering{  20.3  }  &  \centering{  9.1  }  &  \\        
  &            &        &        &        &        &  \\        
\raggedright \citet{jager94}&  \raggedright Melting and Quenching  &  \centering{dirtysil}$^1$  &  \centering{  9.6  }  &  \centering{  1.4  }  &  \centering{  21.3  }  &  \centering{  10.2  }  &  \\        
\raggedright \citet{mutschke98} &  \raggedright Melting and Quenching  &  \centering{  Gehlenite  }  &  \centering{  10.2  }  &  \centering{  1.2  }  &  \centering{  21.4  }  &  \centering{  7.5  }  &  \\        
  &    &  \centering{  melilite  }  &  \centering{  10.5  }  &  \centering{  1.8  }  &  \centering{  19.9  }  &  \centering{  5.6  }  &  \\        
  &            &        &        &        &        &  \\        
\raggedright \citeauthor{dl84} &  \raggedright synthetic  &  \centering{  generic  }  &  \centering{  10.5  }  &  \centering{  3.3  }  &  \centering{  20.8  }  &  \centering{  8.6  }  &  \\        
\raggedright \citeauthor{ohm92} warm  &  \raggedright synthetic  &  \centering{  generic  }  &  \centering{  10.7  }  &  \centering{  2.5  }  &  \centering{  19.7  }  &  \centering{  5.7  }  &  \\        
\raggedright \citeauthor{ohm92} cold  &  \raggedright synthetic  &  \centering{  generic  }  &  \centering{  10.8  }  &  \centering{  2.6  }  &  \centering{  19.7  }  &  \centering{  5.3  }  &  \\        
\hline
\end{tabular}
\begin{tabular}{p{6.5in}}
$^1$ Mg$_{0.50}$Fe$_{0.43}$Ca$_{0.03}$Al$_{0.04}$SiO$_3$ \\
\end{tabular}
\end{table}

\clearpage
\begin{table}
\caption{Glass Sample Compositions and physical parameters}
\footnotesize
\label{tab:samples}
\begin{tabular}{l@{\hspace{1mm}}l@{\hspace{1mm}}c@{\hspace{1mm}}c@{\hspace{1mm}}c@{\hspace{1mm}}c}
\hline
Name & Formula  & $\rho$ (kg\,m$^{-3}$) & $T_g$ (K)	& Al/(Al$+$Si)& NBO/T$^a$ \\
\hline
Synthetic Gehlenite         & $\rm (Na_{0.01}Ca_{1.99})Al(Al_{1.03}Si_{0.98})O_7 $ & 2879& 1111	&	0.68	& 0.65\\
Gehlenite Remelt  & $\rm Na_{0.02}Ca_{1.97}Mg_{0.37}Fe_{0.06}Al_{1.14}Si_{1.44}O_7$ & 2927 & 1032	&	0.44	& 1.43 \\
Synthetic Akermanite        & $\rm Ca_{2.04}Mg_{0.98}Al_{0.02}Si_{1.97}O_7$ & 2951 &1012	&	0.01	& 3.04\\
Forsterite        & $\rm Mg_{2.035}Si_{0.983}O_4$& 2920 & 1040$^b$	&	0.00	& 4.00\\
Diopside & $\rm Ca_{1.02}Mg_{0.92}Si_{2.03}O_6$ & 2853& 999      & 0.00       &   1.91    \\
Akermanite Remelt & $\rm Na_{0.30}Ca_{1.70}Mg_{0.47}Fe_{0.29}Al_{0.54}Si_{1.64}O_7$ & 2995 & 932	&	0.25	& 2.15\\
Basalt  & $\rm K_{0.01}Na_{0.06}Mg_{0.45}Ca_{0.54}Fe_{0.32}Ti_{0.09}Al_{0.70}Si_{1.72}O_6$   &       &    &   0.03	& 1.48\\
Synthetic Enstatite         & $\rm (Mg_{1.98}Al_{0.02})(Al_{0.01}Si_{1.99})O_6 $ 	& 2735 & 1037	&	0.02	& 1.94\\
Enstatite remelt  & $\rm Ca_{0.02}Mg_{1.81}Fe_{0.14}Al_{0.03}Si_{2.00}O_6$ & 2769 & 1022	&	0.01	& 1.93\\
``Cosmic Silicate''& $\rm (Na_{0.11}Ca_{0.12}Mg_{1.86})(Al_{0.18}Si_{1.85})O_6 $& 2772 & 1001	&	0.09	& 1.92\\
Herasil            & $\rm Si0_2 [>0.1wt\% H_2O]$& 2200$^c$ & 1420	&	0.00	& 0.00\\
Obsidian         & $\rm K_{0.28}Na_{0.29}Ca_{0.02}Fe_{0.04}Al_{0.64}Si_{3.35}0_8$ & &	& 	0.16	& 0.02	\\
\hline
\end{tabular}
\begin{tabular}{p{6.5in}}
$^a$ NBO/T ${\rm = (K + Na + 2Ca + 2Mg + 2Fe^{2+} - Al - Fe^{3+}) / (Si + Ti + Al + Fe^{3+})}$\\
$^b$ from Tangeman et al.\\
$^c$ from manufacturer\\
\end{tabular}
\end{table}

\clearpage

\begin{table*}
\caption{Compositions of glasses, determined by electron microprobe, in weight percent oxides}
\footnotesize
\label{microprobe}
\begin{tabular}{l@{\hspace{2pt}}c@{\hspace{3pt}}c@{\hspace{3pt}}c@{\hspace{3pt}}c@{\hspace{3pt}}c@{\hspace{3pt}}c@{\hspace{3pt}}c@{\hspace{3pt}}c@{\hspace{3pt}}c@{\hspace{3pt}}c@{\hspace{3pt}}c@{\hspace{3pt}}c}
\hline 
	&	\multicolumn{10}{c}{weight percent oxide} & & Water content$^a$\\
name	&	  SiO$_2$& TiO$_2$ & Al$_2$O$_3$ & FeO & MnO & MgO & CaO & Na$_2$O & K$_2$O & P$_2$O$_5$ &	total &(ppm)	\\
\hline																							
Synthetic gehlenite	&	21.53	&	na	&	37.88	&	na	&	na	&	na	&	40.85	&	0.12	&	na	&	na	&	100.37 & 76	\\
Gehlenite remelt	&	31.46	&	0.01	&	21.11	&	1.58	&	0.01	&	5.47	&	40.26	&	0.27	&	0.02	&	0.04	&	100.22	& 85\\
Synthetic \r{a}kermanite	&	43.12	&	na	&	0.44	&	na	&	na	&	14.33	&	41.62	&	na	&	na	&	na	&	99.51 & 80	\\
Forsterite	&	41.17	&	na	&	na	&	na	&	na	&	57.19	&	na	&	na	&	na	&	na	&	98.36	& bd\\
Diopside	&	54.82	&	na	&	na	&	na	&	na	&	16.55	&	25.58	&	na	&	na	&	na	&	96.95	& 57\\
\r{A}kermanite remelt	&	34.34	&	0.89	&	9.65	&	6.85	&	0.52	&	6.53	&	33.32	&	3.08	&	0.49	&	2.03	&	97.72 & 125	\\
Basalt	&	45.92	&	3.04	&	15.99	&	10.16	&	0.13	&	8.04	&	13.45	&	0.80	&	0.18	&	na	&	97.71	& bd \\
Synthetic enstatite	&	59.69	&	na	&	0.68	&	na	&	na	&	39.88	&	na	&	na	&	na	&	na	&	100.25 & 80 	\\
Enstatite remelt	&	58.30	&	0.02	&	0.66	&	4.74	&	0.12	&	35.43	&	0.54	&	0.02	&	0.01	&	na	&	99.85 & 63	\\
Cosmic silicate	&	54.26	&	na	&	4.34	&	na	&	na	&	36.58	&	3.27	&	1.61	&	na	&	na	&	100.05 & 81	\\
Obsidian	&	74.82	&	0.03	&	12.09	&	0.95	&	0.03	&	0.01	&	0.39	&	3.39	&	4.91	&	na	&	96.62 &	$\sim$3000\\
Herasil$^b$	&	100.00	&	na	&	na	&	na	&	na	&	na	&	na	&	na	&	na	&	na	&	100.00 & 940	\\
\hline
\end{tabular}
\begin{tabular}{p{6.5in}}
na = not analyzed (mostly for synthetic melts where the ingredients were well known)	\\
bd = below detection limit \\
$^a$ determined from near-IR spectra using the method in Hofmeister et al. (2009)  \\
$^b$ pure SiO$_2$ with 15-26 ppm metals (Hofmeister and Whittington, in review)\\														
\end{tabular}
\end{table*}

\clearpage
\begin{table}
\caption{Spectral parameters of sample glasses. In all cases the parameters are measured from the absorbance data and the $Q_{\rm abs}$ data 
as plotted in Fig.~\ref{newlabdatafig} and Fig.~\ref{newlabdatafig2}, respectively}
\footnotesize
\label{tab:samples3}
\begin{tabular}{lcccc|cccc|cc|r}
\hline
SAMPLE		&	\multicolumn{4}{c}{$\sim$10\,$\mu$m}				& \multicolumn{4}{c}{$\sim$18\,$\mu$m}				&\multicolumn{2}{c}{10/18\,$\mu$m } & \\
NAME		&	 Peak 	& Barycenter	& \multicolumn{2}{c}{FHWM}	&	 Peak 	& Barycenter	& \multicolumn{2}{c}{FHWM}	&\multicolumn{2}{c}{strength ratio$^\ast$} & Offset$^\dag$\\
		&( $\mu$m)	& ($\mu$m)	&	$a$	&	$Q$	& ($\mu$m)	& ($\mu$m)      &      $a$	&	$Q$	&	$a$	&	$Q$	&\\
Gehlenite	&	10.3	&	10.8	&	2.71	&	2.86	&	       	&		&		&		&		&		&$+$3.25\\
Gehlenite RM	&	10.8	&	10.6	&	2.69	&	2.88	& ---$^\ddag$  	& 	20.0	&	4.08	&	4.45	&	3.74	&	4.25	&$+$2.75\\
Akermanite      &	10.3	&	10.6	&	2.31	&	2.46	&		&		&		&		&		&		&$+$2.50\\
Forsterite	&	10.2	&	10.4	&	2.25	&	2.43	&	       	&		&		&		&		&		&$+$2.25\\
Diopside       	&	10.1	&	10.2	&	2.33	&	2.46	&	19.2	&	19.4 	&	4.34	&	4.12	&	1.65	&	 1.26	&$+$1.50\\
Akermanite RM	&	10.3	&	10.1	&	2.63	&	2.70	&	21.0   	&	20.9	&	4.45	&	4.70	&	1.78	&	1.57	&$+$1.25\\
Basalt         	&	10.0	&	9.9	&	2.27	&	2.41	&	       	&		&		&		&		&		&$+$1.00\\
Enstatite	&	9.9	&	9.9	&	2.38	&	2.58	&	17.6	&	18.5	&	3.99	&	4.70	&	3.03	&	2.25	&$+$0.75\\
Entatite remelt&	10.0	&	9.8	&	2.33	&	2.48	&	        &		&		&		&		&		&$+$0.50\\
Cosmic silicate	&	9.8	&	9.7	&	2.76	&	3.15	&	18.3	&	18.7 	&	4.64	&	4.54	&	4.87	&	4.79	&$+$0.25\\
Obsidian	&	9.0	&	9.1	&	1.91	&	2.14	&	       	&		&		&		&		&		&0.00\\
Herasil		&	9.0	&	8.9	&	1.24	&	1.34	&		&		&		&		&		&		&$-$0.5\\
\hline
\end{tabular}
\begin{tabular}{p{6.5in}}
$^\ast$ strength is defined as the equivalent width (EW) of the feature, rather than the peak-to-continuum ratio. 
Therefore the strength ratio is EW$_{10\mu \rm m}$/ EW$_{18\mu \rm m}$. \\
Peak position for the Gehlenite remelt sample cannot be accurately measured because of interference fringes.\\
$^\dag$ the offset is added to the normalized $Q$-values and is applied in Figure~\ref{newlabdatafig2} for clarity.\\
$^\ddag$ strong fringing prevents measurement of peak position.\\
RM designates remelted samples (see \S~\ref{synth})\\
\end{tabular}
\end{table}


\begin{thebibliography}{}
%

\bibitem[Armstrong(1995)]{Armstrong1995} Armstrong,  J.T.  
1995, Microbeam Anal, 4, 177. 


\bibitem[Bouwman et al.(2001)]{bouwman01}Bouwman, J., Meeus, G., de Koter, A., Hony, S., Dominik, C., Waters, L. B. F. M.  
2001, \aap, 375, 950. 


\bibitem[Casassus et al.(2001)]{casassus01}Casassus, S., Roche, P. F., Aitken, D. K., Smith, C. H. 
2001, \mnras, 320, 424. 

\bibitem[Chiar \& Tielens(2006)]{chiar06}Chiar, J. E., Tielens, A. G. G. M.  
2006, \apj, 637, 774. 

\bibitem[Chiar et al.(2007)]{chiar07}Chiar, J. E., Ennico, K., Pendleton, Y. J., et al. 
2007, \apj, 666, L73. 

\bibitem[Chihara et al.(2007)]{chihara07}Chihara, H., Koike, C., Tsuchiyama, A. 
2007, \aap, 464, 229. 



\bibitem[Day(1979)]{day79}Day, K.L.  
1979, \apj, 234,  158.


\bibitem[Demyk et al.(2000)]{Demyk2000} Demyk, K., Dartois, E., Wiesemeyer, 
H., et al. 2000, \aap, 364, 170.

\bibitem[Demyk et al.(2004)]{Demyk2004} Demyk, K., d'Hendecourt, L., Leroux, H., Jones, A. P., Borg, J. 2004, \aap, 420, 233.

\bibitem[DePew et al.(2006)]{DePew06} 
DePew, K., Speck, A., Dijkstra, C., 2006, \apj, 640, 971.


\bibitem[Dingwell et al.(1996)]{dingwell96}Dingwell, D.B., Romano, C., Hess, K.-U., 
1996, Contributions to Mineralogy and Petrology, 124, 19.

\bibitem[Dorschner et al.(1995)]{dorschner95}Dorschner, J., Begemann, B., Henning, T., Jaeger, C., Mutschke, H. 
1995, \aap, 300, 503. 

\bibitem[Draine(2003)]{draine03} Draine, B.T., 2003, ARA\&A, 41, 241

\bibitem[DL(1984)Draine \& Lee]{dl84}Draine, B. T., Lee, H. M. 
1984, \apj, 285, 89. 

\bibitem[Dwek et al.(2007)]{dwek07}Dwek, Eli, Galliano, Fr�d�ric, Jones, Anthony P.  
2007, \apj, 662, 927. 




\bibitem[Fox(2002)]{fox02}Fox, M. 2002 
``Optical Properties of Solids''
Oxford University Press.

\bibitem[Gail \& Sedlmayr(1999)]{gs99}Gail, H.-P., Sedlmayr, E. 
1999, \aap, 347, 594

\bibitem[Gaustad(1963)]{gaustad63}Gaustad, John E. 
1963, \apj, 138, 1050. 


\bibitem[Getson \& Whittington(2007)]{getson07}Getson, J.M., Whittington, A.G. 
2007Journal of Geophysical Research, 112, B10203, 

\bibitem[Gillett et al.(1968)]{gillett68}Gillett, F. C., Low, F. J., Stein, W. A.  
1968, \apj, 154, 677. 

\bibitem[Gilman(1969)]{gilman69}Gilman, R.C. 
1969, \apj, 155, L185. 


\bibitem[Glass(1999)]{glass99}Glass, I.S., 1999, 
``Handbook of Infrared Astronomy''
Cambdridge University Press



\bibitem[Grossman(1972)]{grossman72}Grossman, Lawrence 
1972, Geochim. Cosmchim. Acta, 36, 597. 


\bibitem[Hackwell et al.(1970)]{hackwell70}Hackwell, J. A., Gehrz, R. D., Woolf, N. J. 
1970, Nature, 227, 822. 

\bibitem[Hao et al.(2005)]{hao05}Hao, Lei, Spoon, H. W. W., Sloan, G. C., et al. 
2005, \apj, 625, L75. 

\bibitem[Henderson(2005)]{henderson}Henderson, G.S., 2005, 
The Canadian Mineralogist, 43, 1921.

\bibitem[Henning \& Stognienko(1993)]{henning93} Henning, Th., Stognienko, R.
1993 \aap, 280, 609.


\bibitem[Hofmeister et al.(2009)]{hofmeister09}Hofmeister, A. M., Pitman, K. M., Goncharov, A. F., Speck, A. K. 
2009, \apj, 696, 1502.

\bibitem[Hofmeister \& Pitman(2007)]{hp07}Hofmeister, A. M., Pitman, K. M. 
2007, Phys. Chem. Min, 34, 319. 

\bibitem[Hofmeister \& Bowey(2005)]{HB06}Hofmeister, A, M., Bowey, J. E., 2006
\mnras, 367, 577.

Monthly Notices of the Royal Astronomical Society, Volume 367, Issue 2, pp. 577-5



\bibitem[Hofmeister, Keppel \& Speck(2003)]{hofm03} Hofmeister, A.M., Keppel, E., Speck, A.K., 2003, \mnras, 345, 16. 

\bibitem[Hony et al.(2009)]{Hony09} Hony, S., Heras, A. M., Molster, F. J., 
Smolders, K. 
2009, \aap, 501, 609.

\bibitem[Huffman \& Stapp(1973)]{huffman73}Huffman, D. R., Stapp, J. L. 
1973, in  Interstellar Dust and Related Topics. IAU Symp. 52, (Eds) J. Mayo Greenberg and H. C. van de Hulst. Dordrecht, Boston, Reidel, p.297

\bibitem[Jaffe et al.(2004)]{jaffe04}Jaffe, W., Meisenheimer, K., R�ttgering, H. J. A., et al. 
2004, Nature, 429, 47. 

\bibitem[J\"{a}ger et al.(1994)]{jager94}J\"{a}ger, C., Mutschke, H., Begemann, B., Dorschner, J., Henning, Th.  
1994, \aap, 292, 641.

\bibitem[J\"{a}ger et al.(2003)]{jager03}J\"{a}ger, C., Dorschner, J., Mutschke, H., Posch, Th., Henning, Th. 
 2003, \aap, 408, 193. 

\bibitem[Jones \& Merrill(1976)]{jm76}Jones, T. W., Merrill, K. M. 
1976, \apj, 209, 509. 

\bibitem[Kemper et al.(2002)]{Kemper2002}Kemper, F., de Koter, A.,Waters, 
L. B. F. M., Bouwman, J., \& Tielens, A. G. G. M. 2002, \aap, 384, 585.


\bibitem[Knacke et al.(1969)]{knacke69}Knacke, R. F., Gaustad, J. E., Gillett, F. C., Stein, W. A. 
1969, \apj, 155, L189. 

\bibitem[Koike et al.(2003)]{koike03}Koike, C., Chihara, H., Tsuchiyama, A., Suto, H., Sogawa, H., Okuda, H. 
2003, \aap, 399, 1101. 


\bibitem[Kr\"{a}tschmer \& Huffman(1979)]{kh79}Kr\"{a}tschmer, W., Huffman, D. R.  
1979, \apss, 61, 195. 


\bibitem[Krishna-Swamy(2005)]{krishna05} Krishna Swarmy, K. S.,  2005
``Dust in the Universe: Similarities And Differences'',
World Scientific Series in Astronomy and Astrophysics, Vol. 7. 
Singapore: World Scientific Publishing, ISBN 981-256-293-1, 2005, XI + 252 pp.

\bibitem[Krugel(2008)]{krugel08} Krugel, E. 2008,
``An Introduction to the Physics of Interstellar Dust''
Taylor \& Francis Group, LLC, New York, p387.



\bibitem[Li(2007)]{li07}Li, A. 
in The Central Engine of Active Galactic Nuclei, ASP Conference Series, Vol. 373, 2007, (Eds) Luis C. Ho and Jian-Min Wang, p.561.

\bibitem[Li et al.(2008)]{li08}Li, M. P., Shi, Q. J., Li, Aigen 
2008, \mnras, 391, L49. 


\bibitem[Li \& Li(2009)]{lili} Li, M. P.,  Li, Aigen
American Astronomical Society, AAS Meeting \#214, \#402.20


\bibitem[Lodders \& Fegley(1999)]{lf99}Lodders, K.,  Fegley, B., Jr. 
1999, in IAU Symp. 191, Asymptotic Giant Branch Stars, ed. T. Le Bertre, A. Lebre, C. Waelkens (New York: Springer), 279


\bibitem[Mann et al.(2006)]{mann06}Mann, Ingrid, K\"{o}hler, Melanie, Kimura, Hiroshi, Cechowski, Andrzej, Minato, Tetsunori, 
2006, A\&ARev, 13, 159.

\bibitem[Marra et al.(2011)]{icarus}Marra, A.C., Lane, M.D., Orofino, C., Blanco, A., Fonti, S.,
2011, Icarus, 211, 839. 

\bibitem[McClure(2009)]{mcclure08}McClure, M. 
2009, \apjl, 693, L81.



\bibitem[Min et al.(2003)]{Min03}
Min, M., Hovenier, J. W., de Koter, A., 2003, \aap, 404, 35.

\bibitem[Min et al.(2007)]{min07}Min, M., Waters, L. B. F. M., de Koter, A., et al. 
2007, \aap, 462, 667

\bibitem[Monnier et al.(1998)]{monnier98}Monnier, J. D., Geballe, T. R., Danchi, W. C. 
1998, \apj, 502, 833. 

\bibitem[Mutschke et al.(1998)]{mutschke98}Mutschke, H., Begemann, B., Dorschner, J., et al. 
1998, \aap, 333, 188. 

\bibitem[Mysen \& Richet(2005)]{mysen05}Mysen, B.O., Richet, P. 
Silicate Glasses and Melts: Properties and Structure. Developments in Geochemistry 10,
Elsevier, 544 pp.

\bibitem[Nuth \& Donn(1982)]{nd82} Nuth, J. A., III, Donn, B. 
1982, \apj, 257, L103

\bibitem[Nuth \& Hecht(1990)]{NuthHecht} Nuth, J. A., III, Hecht, J. H., 
1990, \apss, 163, 79.

\bibitem[Onaka \& Okada(2003)]{OO03} Onaka, T. Okada, Y.,
2003, \apj, 585, 872.

\bibitem[OHM(1992)Ossenkopf, Henning, \& Mathis]{ohm92} Ossenkopf, V., Henning, Th., \& Mathis, J. S. 
1992, \aap, 261, 567.


\bibitem[Papoular \& P\'{e}gouri\'{e}(1983)]{papoular83}Papoular, R., P\'{e}gouri\'{e}, B. 
1983, \aap, 128, 335.


\bibitem[Perets et al.(2010)]{perets} Perets, H.B.,  Gal-Yam, A.,  and Mazali, P.A., et al. 
2010, Nature, 465, 322.

\bibitem[Pollack et al.(1973)]{pollack73}Pollack, J. B., Toon, O. B., Khare, B. N. 
1973, Icarus, 19, 372. 

\bibitem[Richet et al.(1993)]{richet1993}Richet, P., Leclerc, F., Benoist, L. 
1993, Geophys. Res. Lett., 20, 1675.

\bibitem[Sargent et al.(2010)]{Sargent2010} Sargent, B., Srinivasan, S., Meixner, M., et al., 2010, \apj, 716, 878.

\bibitem[Sargent et al(2006)]{sargent06} Sargent, B., Forrest, W. J., D'Alessio, P, et al., 2006, \apj, 645, 395.

\bibitem[Scott \& Duley(1996)]{sd96}Scott, A., Duley, W. W. 
1996, \apjs, 105, 401. 


\bibitem[Schairer \& Bowen(1956)]{bowen} 
Schairer, J.F., Bowen, N.L. 1956, 
American Journal of Science, 254, 129.



\bibitem[Sloan et al.(2003)]{sloan03}
Sloan, G. C., Kraemer, Kathleen E., Goebel, J. H., Price, S. D., 2003, \apj, 594, 483.

\bibitem[Speck(1998)]{speck98}Speck, A. K.  
1998, PhD Thesis.

\bibitem[Speck et al.(2000)]{speck00}Speck, A. K., Barlow, M. J., Sylvester, R. J., Hofmeister, A. M. 
2000, \aaps, 146, 437. 

\bibitem[Speck \& Hofmeister(2004)]{SH04} Speck, A.K., Hofmeister, A.M., 
2004, \apj, 600, 986.

\bibitem[Speck, Thompson \& Hofmeister(2005)]{STH05} Speck, A.K., Thompson, 
G.D., Hofmeister, A.M., 
2005, ApJ, 634, 426.

\bibitem[Speck et al.(2008)]{swt08}Speck, Angela K., Whittington, Alan G., Tartar, Josh B. 
2008, \apj, 687, L91. 

\bibitem[Speck et al.(2009)]{speck09}Speck, Angela K., Corman, Adrian B., Wakeman, Kristina, Wheeler, Caleb H., Thompson, Grant 
2009, \apj, 691, 1202.


\bibitem[Stencel et al.(1990)]{stencel90}Stencel, Robert E., Nuth, Joseph A., III, Little-Marenin, Irene R., Little, Stephen J. 
1990, \apj, 350, L45

\bibitem[Stolper(1982)]{stolper82}Stolper, E. 
1982, Geochim. Cosmochim. Acta, 46, 2609.

\bibitem[Stroud et al.(2008)]{stroud08}Stroud, R. M., Nguyen, A. N., Alexander, C. M. O'd., Nittler, L. R., Stadermann, F. J. 
2008, Meteoritics \& Planet. Sci, Abst. 43, 5201. 

\bibitem[Sugerman et al.(2006)]{sugerman06}Sugerman, Ben E. K., Ercolano, Barbara, Barlow, M. J., et al. 
2006, Science, 313, 196. 

\bibitem[Sylvester et al.(1998)]{sylvester98}Sylvester, R. J., Kemper, F., Barlow, M. J., et al. 
1999, \aap, 352, 587.

\bibitem[Tangeman et al.(2001)]{tangeman01a}Tangeman, J.A., Phillips, B.L., Navrotsky, A., Weber, J.K.R., Hixson, A.D.,  Key, T.S. 
2001a, Geophys. Res. Letters, 28, 2517.



\bibitem[Tielens(1990)]{tielens90}Tielens, A. G. G. M. 
1990, in From Miras to Planetary Nebulae: Which Path for Stellar Evolution?, ed. M. O. Mennessier A. Omont (Gif-sur-Yvette: Editions Frontieres), 186

\bibitem[Toppani et al.(2006)]{Toppani2006} Toppani, A., Libourel, G., Robert, F., Ghanbaja, J. 2006, Geochim.\ Cosmochim.\ Acta, 70, 5035.

\bibitem[van Breeman et al.(2011)]{vanbreeman}
van Breemen, J. M.; Min, M.; Chiar, J. E.;  et al.
2011, \aap, in press.

\bibitem[Videen \& Kocifaj(2002)]{vk02} Videen, G., Kocifaj, M., 2002
Optics of Cosmic Dust: proceedings of a NATO Advanced Research Workshop
NATO Science Series. 
Dordrecht/Boston/London: Kluwer Academic Publishers, 2002.

\bibitem[Volk \& Kwok(1988)]{vk88} Volk, K., Kwok, S., 
1988, ApJ, 331, 435.


\bibitem[Vollmer et al.(2007)]{vollmer07}Vollmer, C., Stadermann, F. J., Bose, M., Floss, C., Hoppe, P., Brenker, F. E.  
2007 Meteoritics \& Planet. Sci, Abst. 42, 5107. 

\bibitem[Voshchinnikov \& Henning(2008)]{vh08}Voshchinnikov, N. V., Henning, T.  
2008, \aap, 483, L9. 



\bibitem[Watkins et al.(2009)]{watkins09}Watkins, J., Manga, M., Huber, C., Martin, M. 
2009. Contributions to Mineralogy and Petrology, in press, DOI 10.1007/s00410-008-0327-8

\bibitem[Wheeler \& Speck(2007)]{wheeler07}Wheeler, Caleb, Speck, A. K. 
2007, BAAS, 39, 890.

\bibitem[Wheeler et al(2011)]{wheeler09}Wheeler, Caleb, Speck, A. K. 
2011, in prep.

\bibitem[Whittet(1992)]{Whittet1992} Whittet, D. C. B., 1992, ``Dust In The Galactic Environment'', IoP Publishing.

\bibitem[Whittington et al.(2009)]{whittington09}Whittington AG, Hellwig BM, Behrens H, Joachim B, Stechern A, and Vetere F, 
2009, Bulletin of Volcanology, 71, 185.

\bibitem[Wilding et al.(1995)]{wilding95}Wilding, M.C., Webb, S.L., Dingwell, D.B. 
1995. Chemical Geology, 125, 137.



\bibitem[Woitke(2006)]{woitke06}Woitke, P.  
2006, \aap, 460, L9.

\bibitem[Woolf(1973)]{woolf73}Woolf, N. J. 

\bibitem[Woolf \& Ney(1969)]{woolf69}Woolf, N. J., Ney, E. P.  
1969, \apj, 155, L181. 

\bibitem[Zachariasen(1932)]{zachariasen}
Zachariasen, W. H.,  1932, 
Journal of the American Chemical Society, 54, 3841.


%
%

%
%

\end{thebibliography}
\end{document}